\documentclass{aa} 
\usepackage{graphicx}
\usepackage{txfonts}

\usepackage{subfigure}

\newcommand{\herschel}{{\it Herschel}}
\newcommand{\wise}{{\it WISE}}
\newcommand{\um}{\,$\mu$m}

\usepackage{color}

\begin{document}

   \title{Sifting the debris: Patterns in the SNR population with unsupervised ML methods}

   \subtitle{}

   \author{F. Bufano\inst{1} \thanks{Present address: filomena.bufano@inaf.it},
          C. Bordiu\inst{1},       
          T. Cecconello\inst{1}\inst{2}\inst{3},
          M. Munari\inst{1},
          A. Hopkins\inst{4}, 
        A. Ingallinera\inst{1},
        P. Leto\inst{1},
        S. Loru\inst{1},
        S. Riggi\inst{1},
        E.Sciacca\inst{1}, 
        G. Vizzari\inst{3},
        A. De Marco\inst{5},
        C.S. Buemi\inst{1}, 
        F. Cavallaro\inst{1},
        C.Trigilio\inst{1}
        \and
        G. Umana\inst{1}
    }

   \institute{INAF - Osservatorio Astrofisico di Catania, Via Santa Sofia 78, 95123, Catania, Italy
         \and Department of Electrical, Electronic and Computer Engineering, University of Catania, Viale Andrea Doria 6, 95125, Catania, Italy        
         \and Universit\'a degli Studi di Milano-Bicocca, Viale Sarca 336, 20126, Milano, Italy 
         \and School of Mathematical and Physical Sciences, 12 Wally’s Walk, Macquarie University, NSW 2109, Australia
         \and Institute of Space Sciences and Astronomy, Maths \& Physics Building, University of Malta, Msida MSD2080, Malta}

   \date{Received ..; accepted ..}

  \abstract
  % context heading (optional)
  % {} leave it empty if necessary  
   {Supernova remnants (SNRs) carry vast amounts of mechanical and radiative energy that heavily influence the structural, dynamical, and chemical evolution of galaxies. 
To this day, more than 300 SNRs have been discovered in the Milky Way, exhibiting a wide variety of observational features. 
However, existing classification schemes are mainly based on their radio morphology. }
  % aims heading (mandatory)
   {In this work, we introduce a novel unsupervised deep learning pipeline to analyse a representative subsample of the Galactic SNR population ($\sim$ 50\% of the total) with the aim of finding a connection between their multi-wavelength features and their physical properties.}
  % methods heading (mandatory)
   {The pipeline involves two stages: (1) a representation learning stage, consisting of a convolutional autoencoder that feeds on imagery from infrared and radio continuum surveys (WISE 22$\mu$m, Hi-GAL 70 $\mu$m and SMGPS 30 cm) and produces a compact representation in a lower-dimensionality latent space; and (2) a clustering stage that seeks meaningful clusters in the latent space that can be linked to the physical properties of the SNRs and their surroundings.}
  % results heading (mandatory)
   {Our results suggest that this approach, when combined with an intermediate uniform manifold approximation and projection (UMAP) reprojection of the autoencoded embeddings into a more clusterable manifold, enables us to find reliable clusters. 
Despite a large number of sources being classified as outliers, most clusters relate to the presence of distinctive features, such as the distribution of infrared emission, the presence of radio shells and pulsar wind nebulae, and the existence of dust filaments.}
  % conclusions heading (optional), leave it empty if necessary 
   {}

 \titlerunning{Galactic SNRs classification with machine learning} 
 \authorrunning{F.Bufano et al.}

   \maketitle
%

%________________________________________________________________

\section{Introduction}
\label{sec:introduction}

Supernovae (SNe), the catastrophic endpoint of the evolution of (some classes of) stars, play a pivotal role in the evolution of the Universe: releasing energies of the order of $10^{51}$ erg into the interstellar medium (ISM), they are responsible for triggering star formation, by compressing dense molecular clouds, and also for the acceleration of cosmic particles to extraordinarily high energies. 
Moreover, through their debris (named supernova remnants, hereafter SNRs), they contribute to the chemical enrichment of the environment by spreading around their nucleosynthesis yields and those of the progenitor star.

SNRs are the visible manifestation of the interaction between the SN ejecta and the surrounding circumstellar/interstellar medium.
Thus, their appearance strongly depends on the nature of the SN explosion (type and energy), the time since the explosion (age), the progenitor mass-loss history, and the complexity of the surrounding medium. 
It is for this reason that the radiation from the remnants throughout the electromagnetic spectrum shows a wide morphological and dynamic diversity, and SNRs are thus defined as a heterogeneous population (e.g. see \citealt{Dubner2017} and references therein).

To this day, more than 300 of these sources have been identified in our Galaxy (University of Manitoba Catalogue \footnote{http://snrcat.physics.umanitoba.ca/index.php?}, \citealt{Ferrand2012}), with most of them being discovered based on their radio emission \citep{Green2019}.
Radio emission in SNRs is indeed the predominant manifestation because of the ongoing non-thermal processes (synchrotron emission from the relativistic electrons in magnetic fields).
Nevertheless, important information can be provided by the infrared (IR) emission, which mainly comes from dust (either stochastically or thermally heated).

While most studies on SNRs in the literature focus on the analysis of individual SNR emissions, which are very useful for connecting the SNR with the type of SN explosion and hence with the history of the SN progenitor, there have been only a few limited attempts at a systematic multiwavelength study of their properties (e.g., see \citealt{Pinheiro}, \citealt{Chawner20}, \citealt{Lopez2009}, 2011). As a result, it is difficult to discern between SNRs with peculiar environmental conditions and those with more `standard' characteristics.

Traditionally, SNRs are mainly classified based on their morphology in the radio band: four broad categories have been defined, namely `shell' type, `plerionic' (or  `filled-centre'), `plerionic composite', and the more recently added `mixed morphology' (or `thermal composite') \citep{Dubner2017}. Considering just a single band means that only the physical processes involving one single component are taken into account; in the case of the radio band, this is the ionized gas component.
For this reason, here we present a first experimental attempt at classification of SNRs using a statistically significant  sample spanning a wider range of frequencies (from IR to radio), with the aim being to achieve a classification scheme that considers the underlying physical processes acting on different components.

To this end, we used an innovative approach based on a machine learning (ML) technique. 
This work was developed in two main stages: 1) a representation learning stage using a convolutional autoencoder, which takes imagery from IR and radio continuum surveys as input and returns a compact representation in a lower-dimensionality latent space; and 2) a clustering stage that defines groups of SNRs in the latent space that share common features, possibly linked to physical properties. 

In Sect. \ref{sec:related-works}, we stress the importance of ML methods for astronomical studies and present some examples of their application to scientific use cases similar to the one presented here.
In Sect. \ref{sec:used datasets}, we describe the employed datasets and in Sect. \ref{sec:sample} we present the selected sample.
In Sect. \ref{sec:workflow}, we introduce our workflow step by step, and in Sect. \ref{sec:results} we summarize the experiments we carried out, describing our main attempts and the results obtained.
We discuss our results and outline our main conclusions in Sects. \ref{sec:discussion}  and \ref{sec:conclusions}, respectively.

%__________________________________________________________________

\section{Related works}
\label{sec:related-works}

In recent times, we have witnessed a dramatic increase in the size and complexity of astronomical datasets, with the continuous arrival of new observing facilities ---both ground-based and space-borne--- that provide ever-increasing resolving power, sensitivity, and sky coverage, pushing the limits of our knowledge of the Universe. As the astronomical community stepped into the Big Data era, traditional data processing and analysis pipelines soon fell short of the efficient handling of these large datasets. Consequently, astronomers slowly began resorting to data mining and ML techniques to support their investigations and speed up the transformation of data into valuable scientific insights \citep{2010IJMPD..19.1049B}. 
While the adoption of such techniques was somewhat limited in the 1990s and early 2000s, the continuous development and refinement of more powerful deep learning algorithms, along with the steep increase in hardware capabilities, has led to a quick proliferation of astronomy-related ML applications in the last decade. This uptake has had a remarkable impact on fields as diverse as exoplanet detection (\citealt{Shallue2018} and related papers), photometric redshift estimation \citep{Brescia2021}, morphological classification of galaxies
\citep{Dieleman2015} and galactic sources \citep{Sortino}, gravitational lensing studies \citep{Jacobs2017, Lanusse2018}, and transient detection \citep{Bloom2012, Goldstein2015}, to name a few.

In this context, unsupervised methods, namely those trying to make sense of the data without prior knowledge or `labels', have led to significant improvement in our ability to extract information from large datasets, where (1) labelling becomes impractical, or (2) classification schemes are yet to be defined, and thus exploratory approaches are needed. Very broadly speaking, these methods involve two complementary stages: a first stage of feature extraction, which takes the input data and extracts the most relevant information (e.g. projecting the data into a lower-dimensionality space), and a clustering stage, which attempts to group the data based on a certain similarity metric (e.g. Euclidean distance between points in the feature space). This approach stands out by its ability to yield serendipitous discoveries, unveiling hidden patterns within source populations, revealing new classes of sources, and finding rare objects and anomalies (outliers). Unsupervised methods have thus led to promising results in the classification of stars and galaxies, both from spectra \citep{SanchezAlmeida2010,SanchezAlmeida2013,Fraix2021} and imagery \citep{Spindler2021,Cheng2020}, and the classification of light curves of stars \citep{Varon2011, Naul2018, Valenzuela2018} and supernovae \citep{Rubin2016}.

Regarding SNRs, unsupervised methods have mainly been used to identify regions of interest in individual objects using multi-dimensional data (e.g. \citealt{Iwasaki2019}). However, to the best of our knowledge, no previous attempts have been made to study the Galactic SNR population as a whole from an unsupervised perspective. Therefore, this paper constitutes a first step in the application of such methods to this specific science case. Our particular aim is to decipher whether or not there are any underlying patterns in the SNR population that connect their observed multi-wavelength features to their progenitor stars and their surroundings.

\section{Used datasets}
\label{sec:used datasets}

The shape and brightness  of SNRs at different bands may provide insights that can be used to constrain the physical characteristics of the progenitor stars, their explosion mechanisms, and the ejecta--ISM interaction dynamics. We assembled an image dataset from publicly available radio and far-infrared (FIR) surveys, which together build a comprehensive picture of these sources and their surroundings: the radio, not affected by ISM absorption, provides an unbiased view of the ionised gas emission component; whereas the FIR bands, in particular 22\um\ and 70\um, are well-known tracers of dust grains in SNRs at different temperatures (the longer the wavelength, the lower the temperature).

We employed the following surveys:
 \begin{itemize}
     \item WISE  (Wide-Field Infrared Survey Explorer, \citealt{wise}), an all-sky survey in four IR bands, namely 3.4\um , 4.6\um , 12\um,\,  and 22\um\ \footnote{Data used in this work correspond to the March 14, 2012 release, \url{https://wise2.ipac.caltech.edu/docs/release/allsky/}}. We used exclusively the 22\um\ images, with a native angular resolution of 12 arcsec and a 5$\sigma$ point source sensitivity of better than 6 mJy. \\
     
    \item Hi-GAL (Herschel infrared Galactic Plane Survey, \citealt{HIGAL}), a Galactic Plane survey performed using the Photoconductor Array Camera and Spectrometer (PACS; \citealt{PACS}) and the Spectral and Photometric Imaging Receiver (SPIRE; \citealt{SPIRE}) instruments on board the \herschel\, Space Observatory \citep{pilbratt}. Hi-GAL mapped the inner part of the Galaxy ($|l| \leq 70^\circ, \quad |b| \leq 1^\circ$) in five wavebands, namely  70\um, 160\um, 250\um, 350\um,\,  and 500\um, providing well-sampled coverage of the wavelength range within which the spectral energy distribution of cold dust peaks. In this work, we used only the maps at 70\um\ with a native angular resolution of 6.7 arcsec.\\
    
    \item The SMGPS (SARAO MeerKAT Galactic Plane Survey, \citealt{Goedhart}), the deepest radio continuum survey in L-band to date,  which covers a large portion of the first, third, and fourth Galactic quadrants ($l$=$2^{\circ}-60^{\circ}$, $252^{\circ}-358^{\circ}$, $|b|<$1.5$^{\circ}$) in the frequency range 886-1678 MHz \citep{Goedhart}. For this work, we used the total intensity maps made publicly available in the first data release. These maps were produced by fitting all 14 of the frequency channels and giving the flux density at the 1284 MHz reference frequency. The average rms outside the Galactic plane  for a point-like source is of  the order of $\sim$30 $\mu$Jy beam$^{-1}$, with a synthesized beam of $\sim$8$\times$8 arcsec$^{2}$.  \\
\end{itemize}

The technical details of these  surveys are summarized in Table \ref{tab:survey-tech}

\begin{table*}
\caption{Technical details of the IR and radio surveys used in this work.}
\label{tab:survey-tech}
\begin{tabular}{@{}llccllll@{}}
\hline
Survey         & $\lambda$ & FWHM      & Pixel size & Coverage   &  Unit   & Projection     & Reference \\ 
                &           & (arcsec)  & (deg)                &            &         &         &        \\ 
\hline
WISE            & 22 $\mu$m & 12        & $3.3\times10^{-4}$   & all sky    &   DN    & SIN      & \cite{wise}             \\
Hi-GAL           & 70 $\mu$m & 6.7       & $8.8\times10^{-4}$   & $|l| \leq 70^\circ, \quad |b| \leq 1^\circ$ & MJy sr$^{-1}$ & TAN  & \cite{HIGAL}       \\
SMGPS           & 30 cm     & 8         & $4.2\times10^{-4}$   & $l$=$2^{\circ}-60^{\circ}$, $250^{\circ}-358^{\circ}$, \quad $|b|\leq$1.5$^{\circ}$ &Jy beam$^{-1}$  & SIN   & \cite{Goedhart}   \\ 
\hline
\end{tabular}
\end{table*}

\section{Sample selection}
\label{sec:sample}

The most up-to-date census of Galactic SNRs consists of 383 objects, according to the SNR catalogue by the University of Manitoba \footnote{http://snrcat.physics.umanitoba.ca/index.php?} \citep{Ferrand2012}. Of these, 294 correspond to the Radio Catalogue of Galactic Supernova Remnants by \cite{Green2019}, and the remaining 89 are candidates or confirmed objects from other works. As the starting point for our sample, we took the entire \cite{Green2019} catalogue plus candidates from the MOST Supernova Remnant Catalogue (MSC.C\footnote{\url{http://www.physics.usyd.edu.au/astrop/wg96cat/msc.c.html}}). The main constraint for building a representative sample from this list was the availability of imagery, that is, the sky coverage of the employed surveys. In this respect, Hi-GAL and MeerKAT were the most stringent (see Table \ref{tab:survey-tech}), reducing the number of sources to 223.

For each of these, we produced square cutouts using a custom wrapper of the \textsc{montage}\footnote{http://montage.ipac.caltech.edu/} code, employing mosaicking in those cases where a source fell in between two adjacent survey tiles. 
The cutouts were taken larger than the size of the SNRs, with a side two times the corresponding SNR radius reported in the literature. We then manually assessed the quality of the resulting cutouts, excluding those sources that satisfy at least one of the following conditions:
\begin{itemize}
    \item unresolved or too compact to provide any useful morphological information;
    \item affected by strong imaging artefacts in any of the three bands;
    \item located in packed regions of the Galactic plane, where confusion with unrelated extended sources (e.g. nearby or overlapping \ion{H}{ii} regions) may introduce biases;
    \item located excessively close to the survey coverage edges in any of the three bands, thus resulting in cropped images.
\end{itemize}

\begin{figure*}
    \centering
    \includegraphics[width=1.1\textwidth]{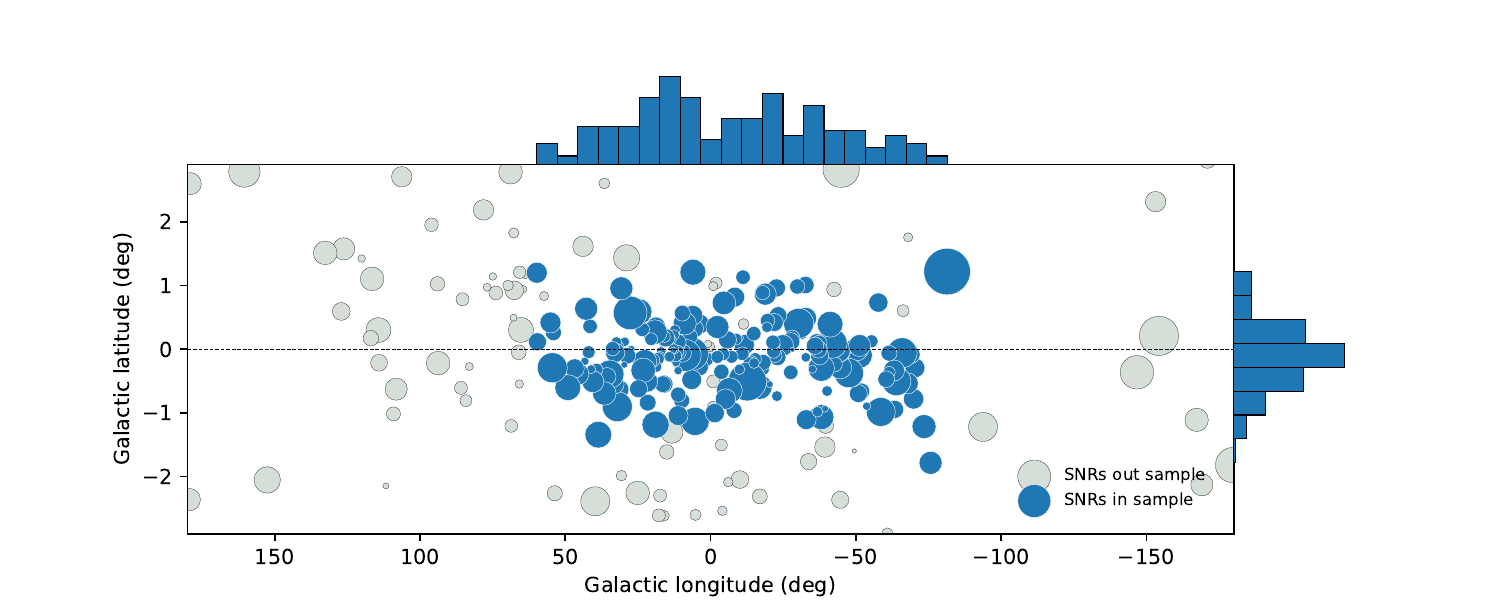}
    \caption{Distribution of Galactic SNRs. The SNRs included in this work, meeting the criteria described in Sect. \ref{sec:sample}, are shown in blue. The other SNRs in \citealt{Ferrand2012} are in grey. Dot size is proportional to SNR radius. The dashed line indicates $b=0$ deg. The top and right histograms represent the marginal distributions of the sample in Galactic longitude and latitude.}
    \label{fig:sample-distribution}
\end{figure*}

The application of these criteria cut down the sample to 178 objects, representing nearly $\sim50\%$ of the total population. Figure \ref{fig:sample-distribution} shows the $l, b$ distribution of the sources that constitute the final sample. A complete list of the sources is provided in Table \ref{Tab:SNR_list}.

\section{Unsupervised workflow for classification}
 \label{sec:workflow}
 
In this section, we introduce an unsupervised analysis pipeline for classification of SNRs. The overall workflow comprises two steps: a feature-extraction stage, for which we employed a convolutional autoencoder (hereafter CAE) to scan the input three-channel images, extracting their most relevant features and thus obtaining a compressed representation in a lower-dimensionality latent space; and a clustering stage, where we take the latent vectors, that is, the compressed representations of the input images, and explore several clustering strategies to find physically meaningful groups. 
In the following, a more detailed description of each step is given. 

\subsection{Dataset preparation}
\label{subsec:dataprep}

As explained in Section \ref{sec:sample}, the input dataset consists of 178 sources, each of them represented by a three-channel RGB image (R=SMGPS, G=Hi-GAL, B=WISE) containing the desired multi-wavelength morphological information needed to classify the SNR population. 
Raw cutouts cannot be directly fed into the CAE, as autoencoders typically require that input images have the same dimensions and be properly normalized. In our case, each cutout has a different size, which is proportional to the source angular radius. Furthermore, for a given source, each channel has a different number of pixels, angular resolution, and brightness units owing to the technical differences between the surveys (see Table \ref{tab:survey-tech}). To overcome these issues and homogenise the sample, a number of preprocessing steps are required:\\

\noindent {\bf Convolution}.
The first step involves, for each source, the convolution of the three channels to a common beam in order to obtain a common angular resolution and thus make the features in the different bands directly comparable. To compute the common beam, we used the Khachiyan algorithm \citep{khachiyan} as implemented in the \texttt{radiobeam} Python package. 
In our case, the common beam turns out to be equal to that of WISE (12 arcsec). As the beams are all circular and not drastically different in size, the net effect of the convolution is a subtle smoothing of the Hi-GAL and SMGPS images.\\

\noindent
{\bf Reprojection and regridding}.
After convolution, we had to correct for the different projection systems of each survey (see Table \ref{tab:survey-tech}) in order to obtain properly aligned channels. 
This was done using the \texttt{reproject} task of the \textsc{Montage} package, which also allows  regridding of the channels to a common pixel grid, producing channels with the same number of pixels. Finally, we conveniently regridded the resulting channels to a fixed size of $64\times64$ pixels, obtaining 178 RGB images of equal size ($3\times64\times64$) for the CAE. 
This value was found to be a good compromise solution between the number of pixels ---which is critical for the model performance--- and the level of detail retained in the images (see Sect.\ref{subsec:cae}).\\

\noindent {\bf Flux conversion}
To make all the channels directly comparable from a physical point of view, the pixel intensities must represent the same physical information. 
Therefore, we scaled all the channels appropriately from their native brightness units (see Table \ref{tab:survey-tech}) to a Jy px$^{-1}$ scale \footnote{The WISE data, in units of DN, was applied a DN/Jy conversion factor equal to  5.2269 $\times$ 10$^{-5}$ (see \wise\, explanatory supplement at \url{http://wise2.ipac.caltech.edu/docs/release/allsky/expsup/sec2\_ 3f.html}). \\}.\\

\noindent {\bf Normalisation}.
Convolutional autoencoders perform better if all the input channels are normalised to a common range of values, typically [0,1], especially if they initially have very different dynamic ranges (as is the case here). However, in our case, such normalisation hampers the ability to retain physical information that could be of interest. For instance, variations in the IR-to-radio ratio across the source may convey key information about the processes at work (e.g. thermal/non-thermal), while providing clues as to the relative contribution of warm dust and ionised gas in shocked regions. Therefore, we explored different normalisation strategies. At first, we tried to normalise each image to the maximum and minimum value of all the channels, that is, to the absolute maximum and minimum values in each RGB `cube'. While this allowed us, in principle, to preserve the channel-to-channel intensity ratios, we note that: (1) the cubes exhibit a remarkably broad dynamic range in brightness, with IR channels (WISE and Hi-GAL) typically having average and median values several orders of magnitude higher than the SMGPS channel ---we found that this approach affected the learning process, effectively introducing an undesired bias in the CAE towards the IR features (most prominent in the normalised cubes); and (2) many SNRs exhibit `negative bowls' around the brightest regions in the SMGPS images. Because of the lack of zero-baseline observations, MeerKAT $uv-$coverage is unable to properly sample bright extended emission. This is a known interferometric problem ---well described by \cite{Goedhart}--- that affects the published data products and leads to unphysical IR-to-radio ratios.
Considering these issues, we decided to take a simpler approach, performing a min--max normalisation per channel, and acknowledging this as a major limitation of the work.\\

\noindent {\bf Compact source removal}.
\label{subsubsec:additional_cons}
All the cutouts are populated by a varying number of compact sources, mostly stellar objects or background galaxies, which are particularly problematic in the WISE 22\um\ band, where the fields are sometimes heavily crowded. These compact sources are not related to the SNRs and thus do not convey useful morphological information for the feature extraction stage. They can be considered random `noise' contaminating the sample, that could negatively affect the learning process of the CAE. To deal with this potential issue, we employed the source-finding tool \textsc{caesar} \citep{Riggi2016, RiggiBordiu2021} to produce binary masks for these compact objects  using default source-finding parameters with a significance threshold of 5$\sigma$, as detailed in \citealt{RiggiUmana2021} (Appendix A). The masks were later employed when computing the loss function of the CAE to avoid considering the contaminated pixels (see Sect. \ref{subsec:training}).\\

 \noindent {\bf Masking}.
 SNRs are usually located in complex regions filled with diffuse background emission and neighbouring or overlapping extended sources, such as H\textsc{ii} regions. This situation is particularly critical in the radio channel because of the broad dynamic range of the SMGPS images.  To force the CAE to focus on the SNR features and prevent it from learning unwanted features from neighbouring, unrelated sources, in each cutout  we masked all pixels outside a circular region of 32-pixel radius centred on the SNR.

\subsection{Feature extraction with CAE}
\label{subsec:cae}

An autoencoder is a specific type of neural network designed for unsupervised learning and dimensionality reduction. The main goal is to replicate its own input, that is, mimicking the identity function, so that for an input, $x^{(i)}$ the output $z^{(i)}\to x^{(i)}$. An autoencoder comprises three components: an encoder, a bottleneck, and a decoder. In the encoder stage, the model tries to summarise the main features that describe the input data, reducing dimensionality and thus learning a compressed or encoded representation of the data. This compressed representation constitutes the bottleneck or latent space, which is the only information available for the decoder stage. The decoder then learns to reconstruct the input from this encoded representation as faithfully as possible ---according to a certain loss function. The training process eventually leads to optimisation of the encoded representations, getting rid of the noise and redundancy present in the input data.

\begin{figure*}
    \centering
    \includegraphics[width=0.9\textwidth]{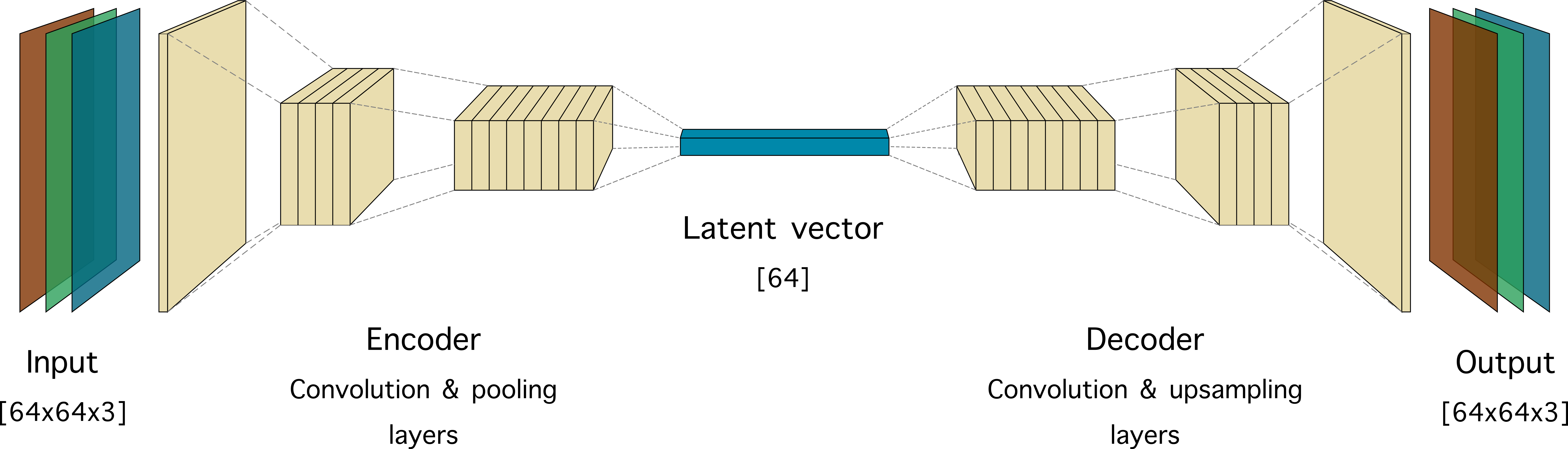}
    \caption{Schematic representation of a convolutional autoencoder architecture, displaying the input (a three-channel RGB image combining WISE, Hi-GAL, and SMGPS imagery), the encoder, the latent vector, the decoder, and the reconstructed output.}
    \label{fig:cae}
\end{figure*}

Convolutional autoencoders represent a variant of the classical autoencoder architecture that is particularly suited for image processing tasks thanks to the use of stacked convolutional and pooling layers. They can scan the input pixels to extract the most relevant features while preserving their spatial relationship. A simple schematic representation of the CAE used in this work is displayed in Figure \ref{fig:cae}. The selected architecture has an encoder stage consisting of three convolutional layers, with a kernel size of (3,3) and an increasing number of filters in each layer; three downsampling layers with a pixel stride of (2,2); and a dense layer of size $D$, which is the dimension of the latent space. All layers employ the Rectified Linear Unit (ReLU) activation function. The decoder stage is simply the mirrored counterpart of the encoder.

In our hyperparameter optimisation process, we concentrated on two key parameters: the number of filters per layer and the dimensionality of the latent space. We considered two flavours of the architecture: the `narrow' architecture (with $32\to64\to128$ filters) and the `wide' architecture (with $64\to128\to256$ filters). The latter was found to perform systematically better in terms of loss (see Sect. \ref{subsec:training}). The selection of the latent space dimensionality $D$ was equally critical, as it affects the quality of the reconstruction (the higher the dimensionality the better) and the `clusterability' of the encoded embeddings (clustering algorithms generally perform better in  spaces of lower dimensionality). We tested several latent space dimensions ranging from 16 to 128 and found 64 to be a reasonable trade-off between image fidelity and clusterability. Therefore, the results presented in the following sections refer to the wide architecture with latent space dimensionality of 64 (unless stated otherwise).

\subsection{Density-based clustering}
\label{subsec:clustering}

The autoencoded embeddings that represent each of the input images in the CAE bottleneck are passed to a clustering stage. While one could rely on the morphological descriptors of SNRs found in the literature (i.e. shell-like, plerionic, composite) as an initial guess, we do not have any prior knowledge about the distribution of the data, and so we favoured an exploratory approach over partition-based clustering schemes, which require that a number of clusters be fixed (like $k-$means). Therefore, our clustering stage employs the density-based spatial clustering of applications with noise (DBSCAN) algorithm first introduced by \cite{ester1996densitybased}. DBSCAN can find clusters of arbitrary shapes and sizes, taking into account the noise in the dataset, or more specifically the outliers, points in the multi-dimensional space that do not necessarily belong to any cluster. The ability to deal with outliers is a key advantage for this work, as one cannot realistically expect every single SNR in the sample to perfectly fit into a certain cluster, considering the heterogeneity of the population.

As we are looking for physically meaningful clusters, the assessment of the clustering outcomes has an inherent subjective component, which is derived from the knowledge of the scientist regarding the underlying physics of SNRs --and how the physics is correlated to multi-wavelength morphological features. However, we also used two standard metrics to evaluate the quality of the clustering:

\begin{itemize}
    \item The Davies-Bouldin index (DBI; \citealt{DaviesBoulding79}), which provides a measure of the similarity of the clusters, by comparing the inter-cluster distance with the cluster size. DBI values closer to zero are desirable, as they indicate better partitions of the dataset. The DBI is computed as in Eq. \ref{eq:dbi}:
    \begin{equation}
        DBI = \frac{1}{k}\sum_{i=1}^{k}\max_{i\neq j}\frac{c_i+c_j}{d_{ij}}
        \label{eq:dbi}
    ,\end{equation}
    where $c_i$ and $c_j$ are the average distances of each point in clusters $i$ or $j$ to the respective cluster centroids, and $d_{ij}$ is the distance between the centroids of clusters $i$ and $j$.\\
    \item The Calinski-Harabasz index (CHI, \citealt{CalinskiHarabasz1974}), which measures the ratio between the dispersion between clusters and the dispersion within clusters, for all clusters. Higher CHI values mean denser, well-differentiated clusters. The CHI index is computed as described in Eq. \ref{eq:chi}:
    \begin{equation}
        CHI = \frac{tr(B_k)}{tr(W_k)}\times\frac{s_E-k}{k-1}
        \label{eq:chi}
    ,\end{equation}
    where $tr(B_k)$ and $tr(W_k)$ represent the traces of the between-cluster dispersion matrix and the within-cluster dispersion matrix, respectively, $k$ is the number of clusters, and $s_E$ is the size of the dataset.
    
\end{itemize}

\section{Experiments}
\label{sec:results}
In this section, we present our experimental results, as obtained with the pipeline described above.

\subsection{Training and validation}
\label{subsec:training}

The input dataset (178 images) was randomly divided into training and validation subsets using a 95--5\% split. 
While having such a low number of samples to work with is not necessarily a problem, the autoencoder training could greatly benefit from having a larger number of examples to learn from, thus allowing better generalisation. 
Therefore, we increased the number of samples through data augmentation techniques. 
We employed four types of transformations, namely asymmetric vertical and horizontal shifting by 15$\%$ of the image range in each direction; vertical and horizontal flipping; rotation by $\pm15^\circ$; and zooming by a randomly chosen factor of $10-15\%$. 
Operations that cause distortions on the image, such as skewing, were disregarded as they would alter the actual morphological features, possibly leading to spurious results.
Each time the model processed the dataset, that is, at every epoch, each image was augmented with a probability of 0.7 with one or more of these transformations. In this way, we prevent the image reconstruction from relying on minimal peculiar details of the image (such as small artefacts, object sizes, or orientations). Each transformation was applied simultaneously to the three image channels to preserve the multi-wavelength pixel-to-pixel relations. 

\begin{figure}
    \centering
    \includegraphics[width=\columnwidth]{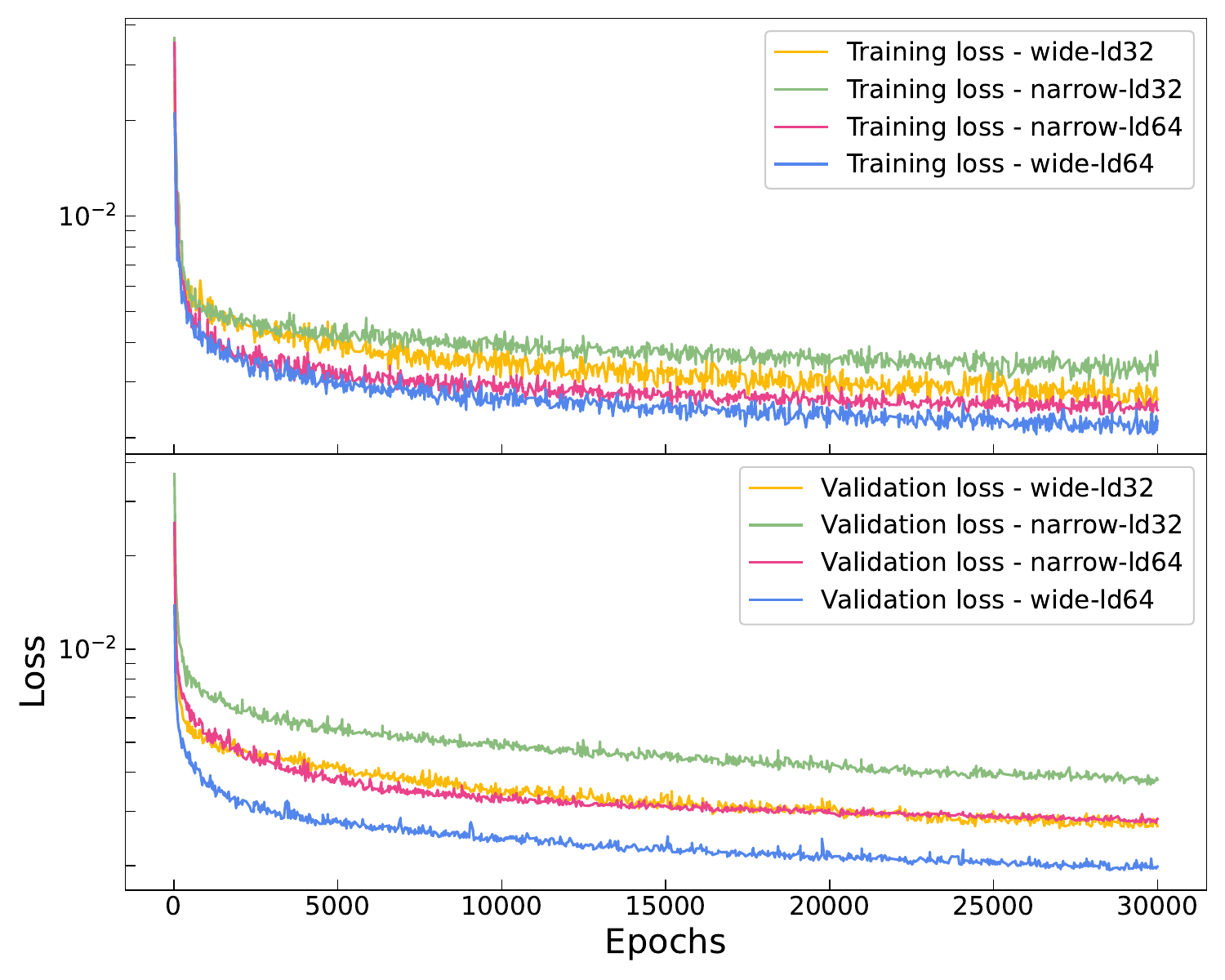}
    \caption{Learning curves (training and validation losses) of four possible CAE setups, combining the \lq narrow\rq\, and \lq wide\rq\, architectures with latent space dimensions of 32 and 64.}
    \label{fig:loss}
\end{figure}

We trained the autoencoder for 30000 epochs (after which the loss was found to stabilise) using the \textit{adam-amsgrad} optimiser \citep{j.2018on}, with a batch size of 64, a learning rate of 0.0001, and a custom mean square error loss function weighted by the masks of the compact sources, as described in Sect.  \ref{subsec:dataprep}. Each training run took $\sim$50 minutes to complete on an NVIDIA 2080 ti GPU. Figure \ref{fig:loss} shows the convergence of the training and validation losses for the different architectures described in Sect. \ref{subsec:cae}. The training and validation losses are systematically lower in the architectures with a 64D latent space, indicating better learning and a better generalisation capability.

\begin{figure*}
    \centering
    \includegraphics[width=\textwidth]{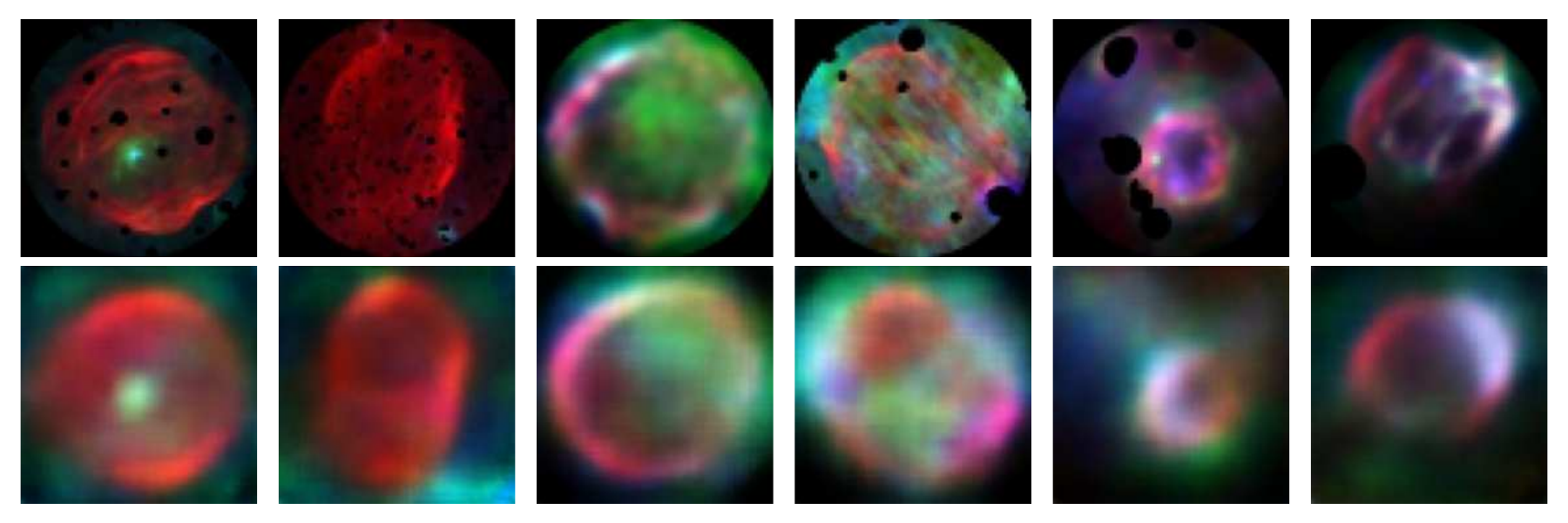}
    \caption{Examples of autoencoder reconstructions. The top row shows a sample of original RGB images, with compact sources masked. The bottom row shows the corresponding reconstructions obtained with the CAE architecture described in Sect. \ref{subsec:cae}}
    \label{fig:reconstructions}
\end{figure*}

In Figure \ref{fig:reconstructions}, we show some examples of the images reconstructed by the autoencoder. We note that, even with a reduced latent space dimension of 64, which implies a compression factor of $\sim200:1$, the CAE does retain the most distinctive features of the input images, while `interpolating' the masked regions (i.e. the compact sources) in a way that provides continuity to the SNR structure.
Finally, the autoencoded embeddings were fed to the clustering stage, where different configurations of hyperparameters were tested in order to look for the best clustering.

\subsection{Clustering the embeddings}
\label{sec:clust_orig}

Our first attempt consisted in clustering the 64-dimension autoencoded embeddings with DBSCAN.
We found the autoencoded embeddings to be remarkably sparse in the latent space, having $\sim$ 180 data points in a 64-dimensional space. 
With such a limited number of data points, the data may be too sparse to contain enough information as to allow for a proper clustering. DBSCAN is extremely sensitive to two complementary hyperparameters: $\epsilon$, the distance between points within a given cluster; and `min\_samples', the minimum number of points required to form a cluster. These parameters need to be properly tuned to avoid all data points being considered outliers or grouped together in a single cluster. Likewise, the selection of an appropriate distance metric is critical. As we are dealing with a latent space, the Euclidean metric, which does not assume any hierarchy in the data features, seems like a reasonable option.

We tested a wide range of hyperparameter values, but DBSCAN struggled to find meaningful clusters. 
In fact, most of the points were classified as outliers by the algorithm, even when setting the min\_samples parameter to low values like 3 to allow a more permissive clustering. This may be an indicator of an inherent lack of structure in the dataset.

\begin{figure}
    \centering
    \includegraphics[width=\columnwidth]{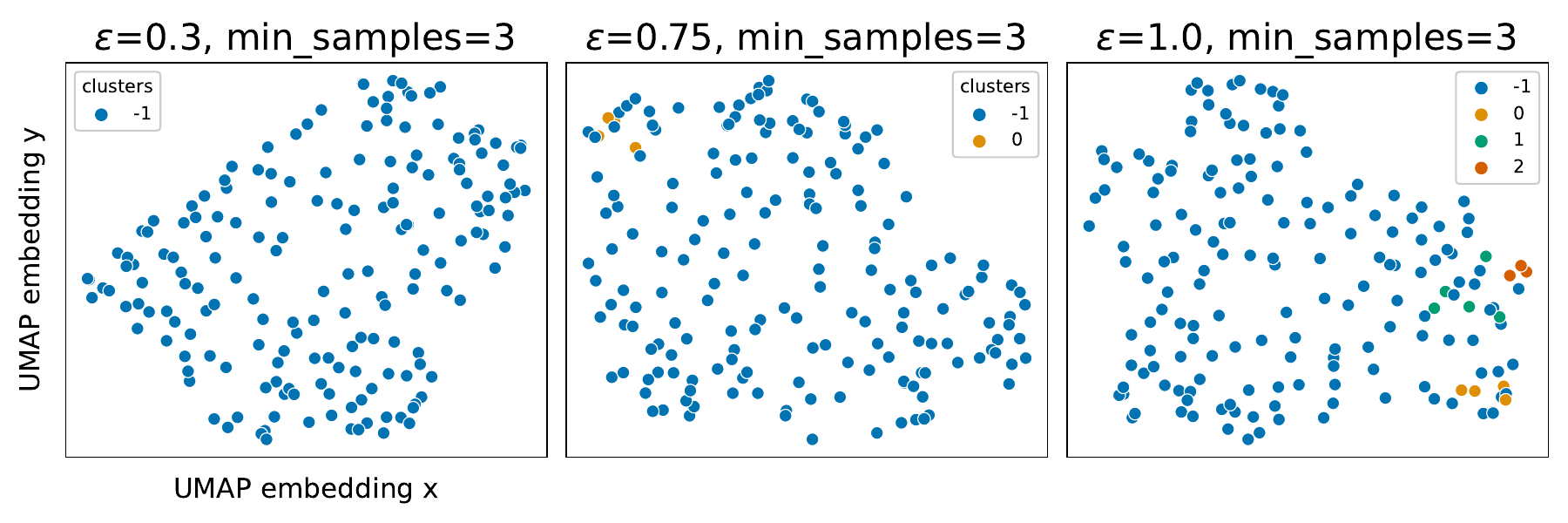}
    \caption{DBSCAN clustering of the original autoencoded embeddings for different hyperparameter values. The 64D embeddings have been projected on a 2D space using UMAP (\textit{n\_neighbors} = 10, \textit{min\_dist} = 0.1). The clustering hyperparameters are displayed at the top of each panel.}
    \label{fig:fig-clust-embed}
\end{figure}

Only small clusters were found even with high $\epsilon$ values, representing less than 10\% of the total number of sources. Figure \ref{fig:fig-clust-embed} shows three clustering examples. 
Only in panel $c$ ($\epsilon$=1) do small clusters appear, seemingly correlated with strong morphological features. For instance, one of the clusters is composed of three sources (\texttt{G054.1+00.3}, \texttt{G292.2-00.5}, \texttt{G318.2+00.1}) that display a shell-like radio component with strong IR emission towards the centre.

\subsection{The N2D approach}
\label{sec:clust_n2d}

The initial clustering results in the original 64-dimensional latent space were unsatisfactory, as only a small fraction of the objects were clustered, on the basis of barely clear common patterns.  
This issue could be related to the `curse of dimensionality', which is a well-known challenge for clustering algorithms when dealing with high-dimensional data. 
In our case, the dataset is clearly high dimensional, with a data points to dimensions ratio of 178:64 $\approx$ 2.8, which translates into a sparse latent space topology where the sources are approximately uniformly distributed.

To overcome this problem and address the sparsity of the latent space in order to improve clusterability, we turned again to dimensionality reduction, using a second feature-extraction stage that takes the autoencoded embeddings as inputs and reprojects them in a space of lower dimensionality. 
In particular, we followed the approach proposed by \cite{McConville2020}, known as `Not 2 (too) deep' clustering, hereafter N2D. 
This non-deep clustering method augments the autoencoder with a manifold learning stage that explicitly takes local structure into account, improving the quality of the representations learned and hence increasing clusterability. 
\cite{McConville2020} found that such a pipeline combined with traditional clustering algorithms matched or outperformed other deep-clustering schemes for a range of benchmark datasets. 
They also concluded that, among other manifold representation methods, UMAP (Uniform Manifold Approximation and Projection; \citealt{umap}) showed the greatest ability to find a clusterable manifold out of the autoencoded embeddings. 
These authors recommend setting the UMAP target dimensionality (the number of UMAP components) equal to a guess of the number of clusters present in the dataset, and the minimum distance between points (min\_dist) to zero, as the primary goal is to obtain an accurate representation of the underlying manifold.

% Please add the following required packages to your document preamble:
% \usepackage{booktabs}
\begin{table*}
\caption{Performed experiments. From left to right: experiment code, dimensionality of the UMAP projection of the autoencoded embeddings (see text), DBSCAN $\epsilon$ value, Calinski-Harabasz index, Davies-Bouldin index, number of clusters, average cluster size, and number of outliers.}
\label{tab:experiments}
\centering
\begin{tabular}{cccccccc}

\hline
Experiment & \#UMAP dims & $\epsilon$ & CHI  & DBI  & \#Cluster & $<\mathrm{Size}>$ & \#Outlier \\ 
\hline
A1         & 4           & 0.30    & 4.15 & 2.27 & 4    & 7.0     & 150       \\
A2         & 4           & 0.35    & 4.58 & 2.30  & 9   & 7.0      & 115       \\
A3         & 4           & 0.40    & 6.58 & 2.59 & 7    & 14.3     & 78        \\
A4         & 4           & 0.45    & 8.79 & 2.98 & 5    & 24.8     & 54        \\\hline
B1         & 6           & 0.30    & 5.03 & 2.06 & 2    & 7.5     & 163       \\
B2         & 6           & 0.35    & 4.70 & 2.50 & 10   & 7.8        & 100       \\
B3         & 6           & 0.40    & 6.39 & 2.65 & 8    & 11.9         & 83        \\
B4         & 6           & 0.45    & 7.79 & 2.84 & 7    & 17.4    & 56        \\\hline
C1         & 8           & 0.30    & 4.99 & 2.10 & 3    & 7.3     & 156       \\
C2         & 8           & 0.35    & 5.00 & 2.36 & 8    & 7.9     & 115       \\
C3         & 8           & 0.40    & 5.95 & 2.73 & 8    & 11.5     & 86        \\
C4         & 8           & 0.45    & 12.2 & 3.28 & 3    & 36.3     & 69        \\\hline
D1         & 10          & 0.30    & 4.68 & 2.20 & 3    & 8.0     & 154       \\
D2         & 10          & 0.35    & 4.66 & 2.43 & 8    & 8.0     & 114       \\
D3         & 10          & 0.40    & 7.06 & 2.83 & 7    & 14.4     & 77        \\
D4         & 10          & 0.45    & 7.58 & 2.83 & 7    & 18.9     & 46        \\\hline
E1         & 12          & 0.30    & 5.54 & 2.11 & 3    & 8.0     & 154       \\
E2         & 12          & 0.35    & 5.16 & 2.51 & 6    & 8.5     & 127       \\
E3         & 12          & 0.40    & 6.29 & 2.65 & 8    & 11.8     & 84        \\
E4         & 12          & 0.45    & 8.24 & 2.86 & 6    & 20.0     & 58        \\ \hline

\end{tabular}
\end{table*}

\begin{figure*}
    \centering
    \includegraphics[width=0.9\textwidth]{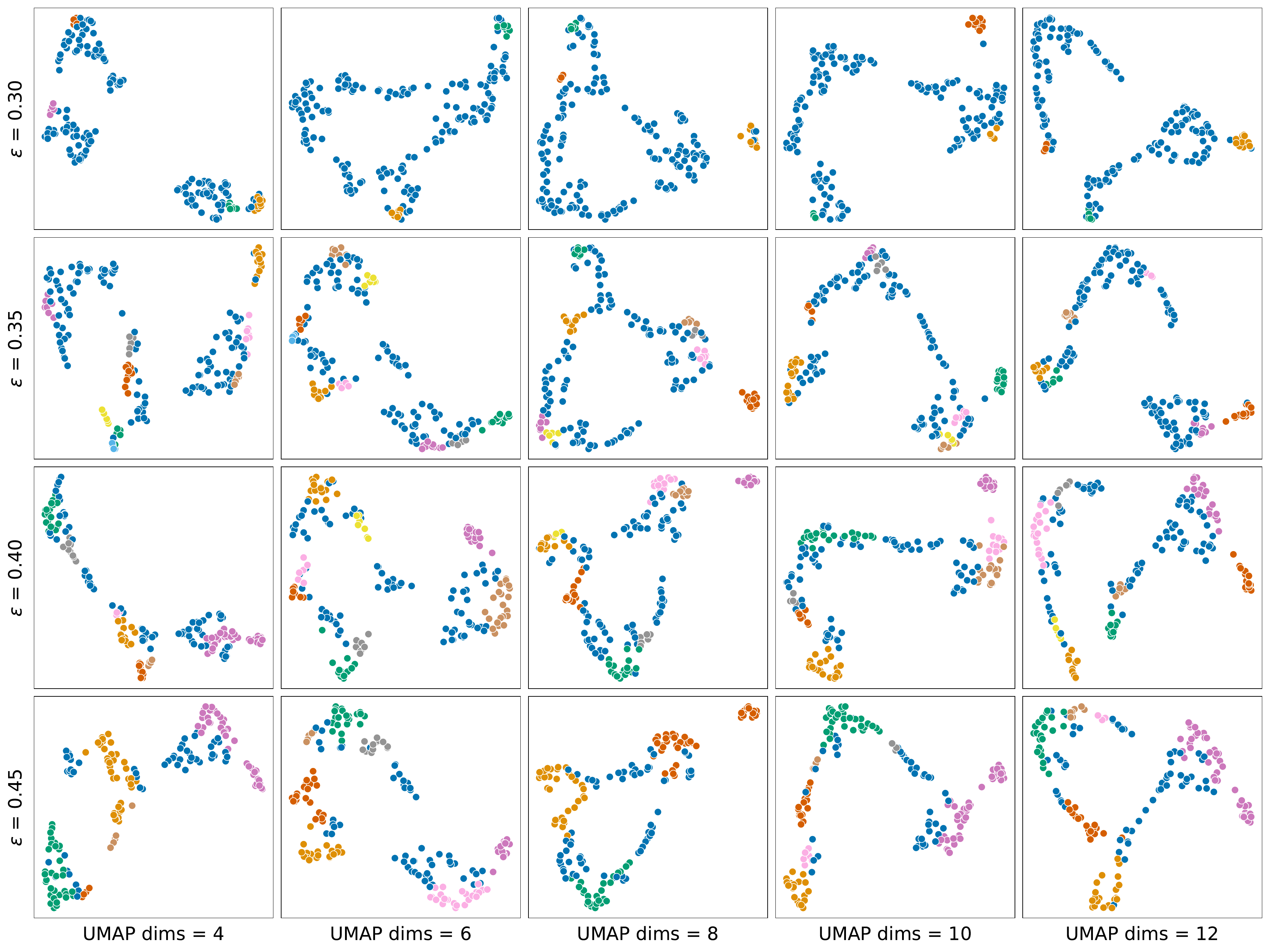}
    \caption{Experiments for different UMAP manifold dimensionality and DBSCAN $\epsilon$ values. }
    \label{fig:experiments}
\end{figure*}

In our case, replicating the N2D pipeline is tricky, as we lack prior knowledge of the number of clusters in the data. As discussed in previous sections, the SNR population is a heterogeneous group for which multiple classification schemes are possible, and so we expect a significant fraction of the data points to be labelled as `outliers', that is, not belonging to any cluster.

For this reason, we employed a blind approach by varying the dimensionality of the projection from 4 to 12 among experiments. We then applied DBSCAN to the resulting manifolds, setting min\_samples to 5 (since smaller clusters may be spurious) and exploring a wide range of $\epsilon$ values. We find that regardless of the dimensionality, the best results in terms of clusterability are achieved with $\epsilon \in [0.3,0.45]$.

Figure \ref{fig:experiments} shows the experiments performed using different dimensionality for the UMAP projection and different DBSCAN values and Table \ref{tab:experiments} lists the number of clusters, average cluster size, number of outliers, and performance metrics for each experiment. 
We note that, while metrics such as CHI and DBI are useful for quantifying the clustering performance, they are not in any way related to the interpretability of the resulting clusters in terms of physical meaningfulness. 
In other words, even spurious clusters may result in good CHI and DBI scores. Therefore, we visually inspected the clustering results to evaluate the intracluster homogeneity, favouring experiments that resulted in strong differences between groups.

\section{Discussion}
\label{sec:discussion}
\subsection{Cluster performance and stability}
In general, we find that the dimensionality of the underlying manifold does not have a decisive impact on the clustering performance. 
For a given UMAP projection, $\epsilon$ values of 0.30 and 0.35 (experiments A1, A2...E1, E2 in Table \ref{tab:experiments}) result in a handful of clusters of small size (less than 10 sources on average) and a large number of outliers (between $60-90\%$ of the sample), a situation not too different from what was obtained by clustering the original 64-dimensional embeddings. 
On the other hand, $\epsilon$ values of 0.45 (experiments A4...E4) tend to produce the lowest number of outliers ($\sim$33\%) but only a few, large clusters. However, having large clusters can severely limit the interpretability of the groups from a physical perspective, as visual inspection reveals that the clusters tend to be more heterogeneous. 
The best compromise between the number and size of the clusters and the number of outliers was achieved for $\epsilon$ = 0.40 (experiments A3...E3 in Table  \ref{tab:experiments}).

\begin{figure}
    \centering
    \includegraphics[width=\columnwidth]{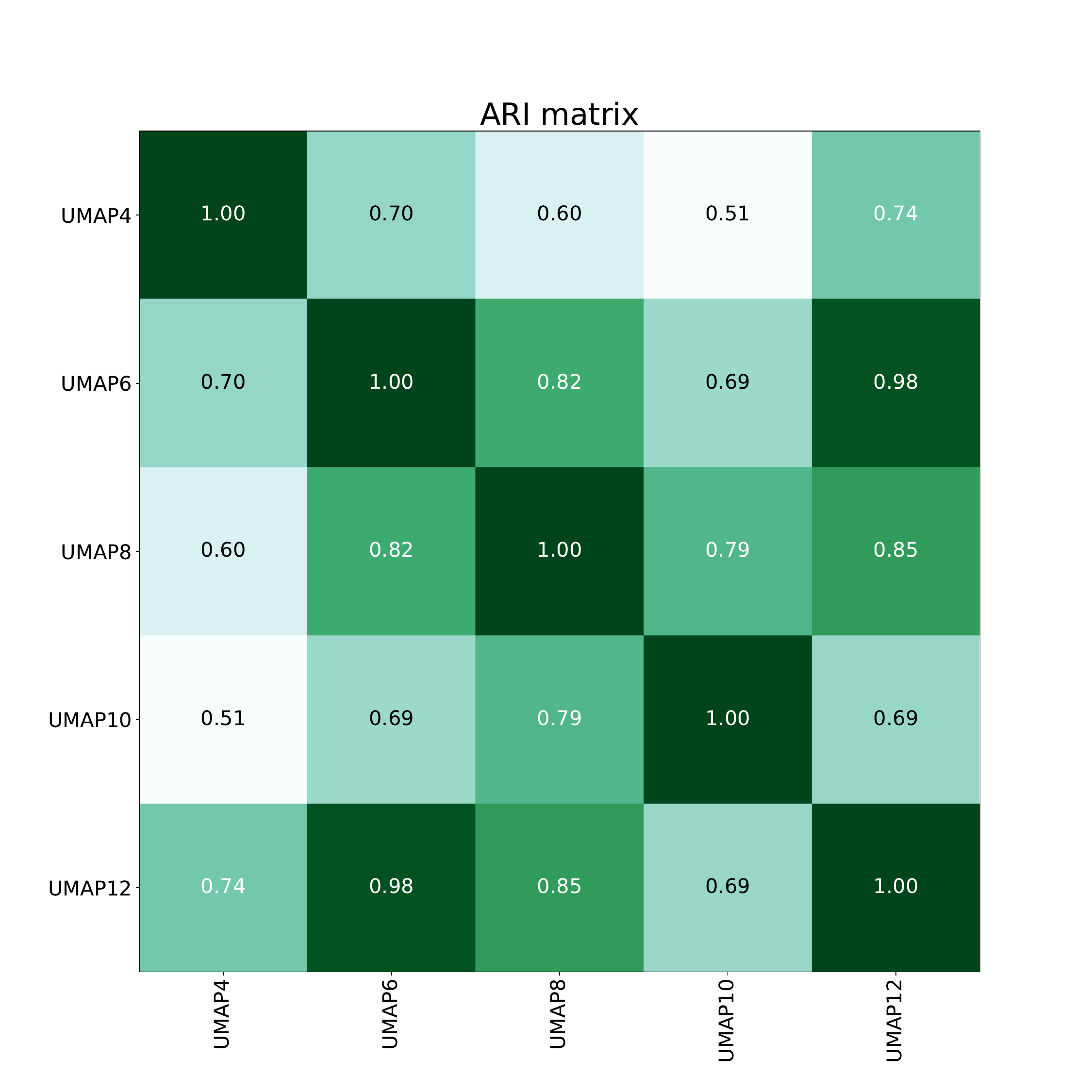}
    \caption{Adjusted Rand index matrix for DBSCAN clusterings with $\epsilon$=0.40}
    \label{fig:ari-matrix}
\end{figure}

To assess the stability of the clusterings with respect to the chosen UMAP dimensionality, we employed the adjusted Rand index (ARI; Halkidi et al., 2002) metric, which is a variant of the Rand index adjusted for chance. 
The ARI measures the similarity between two data partitions (clusterings), approaching 1 when the clusterings are similar, and 0 when the clusterings are random. 
We used ARI to measure the similarity between each pair of experimental results for a fixed DBSCAN parameter configuration. 
However, as the manifold is still relatively sparse, the ARI can be affected by the outliers, especially those points that fall near higher-density regions. For example, a point $P$ may be considered an outlier in experiment $e_i$ but be assigned a label in experiment $e_j$ (depending on the $\epsilon$ value), thus downgrading the ARI metric even if the composition of all the other clusters remains the same. 
To minimise this effect in favour of a `fairer' evaluation of cluster similarity, when computing ARI we only considered points with valid cluster labels in both experiments $e_i$ and $e_j$. 
Figure \ref{fig:ari-matrix} shows an example ARI matrix for experiments A3...E3, with $\epsilon=0.40$. 
The average ARI is $\sim0.74$, but most of the discrepancy involves experiment A3, with UMAP dimensionality of 4. 
At higher dimensionalities, the average agreement increases up to ARI $\sim0.81$, indicating the clusters are stable and therefore not spurious.

\subsection{Cluster physical interpretation}

\begin{figure*}
    \centering
    \subfigure[]{\includegraphics[width=1.2\columnwidth]{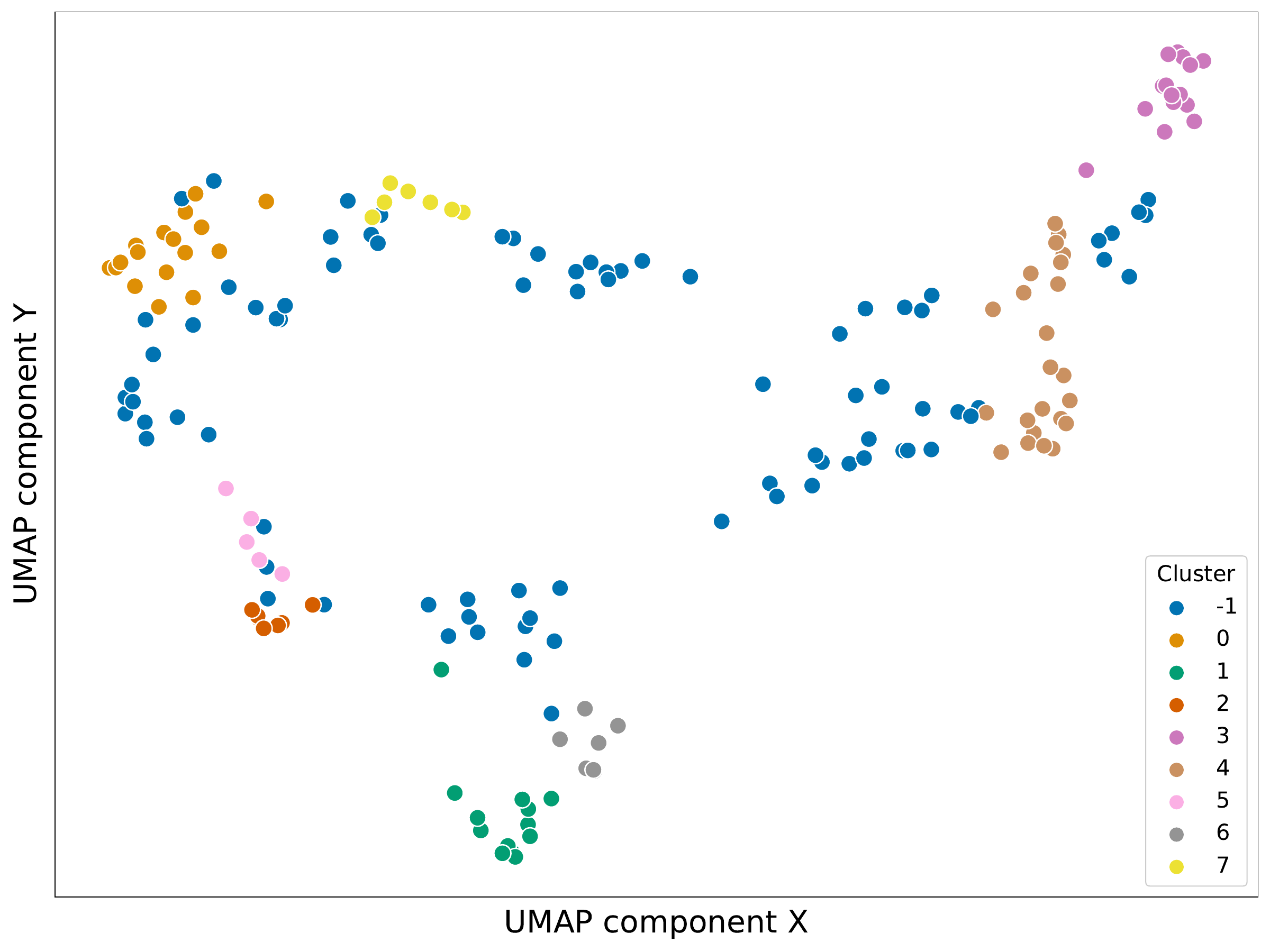} }
    \subfigure[]{\includegraphics[width=1.85\columnwidth]{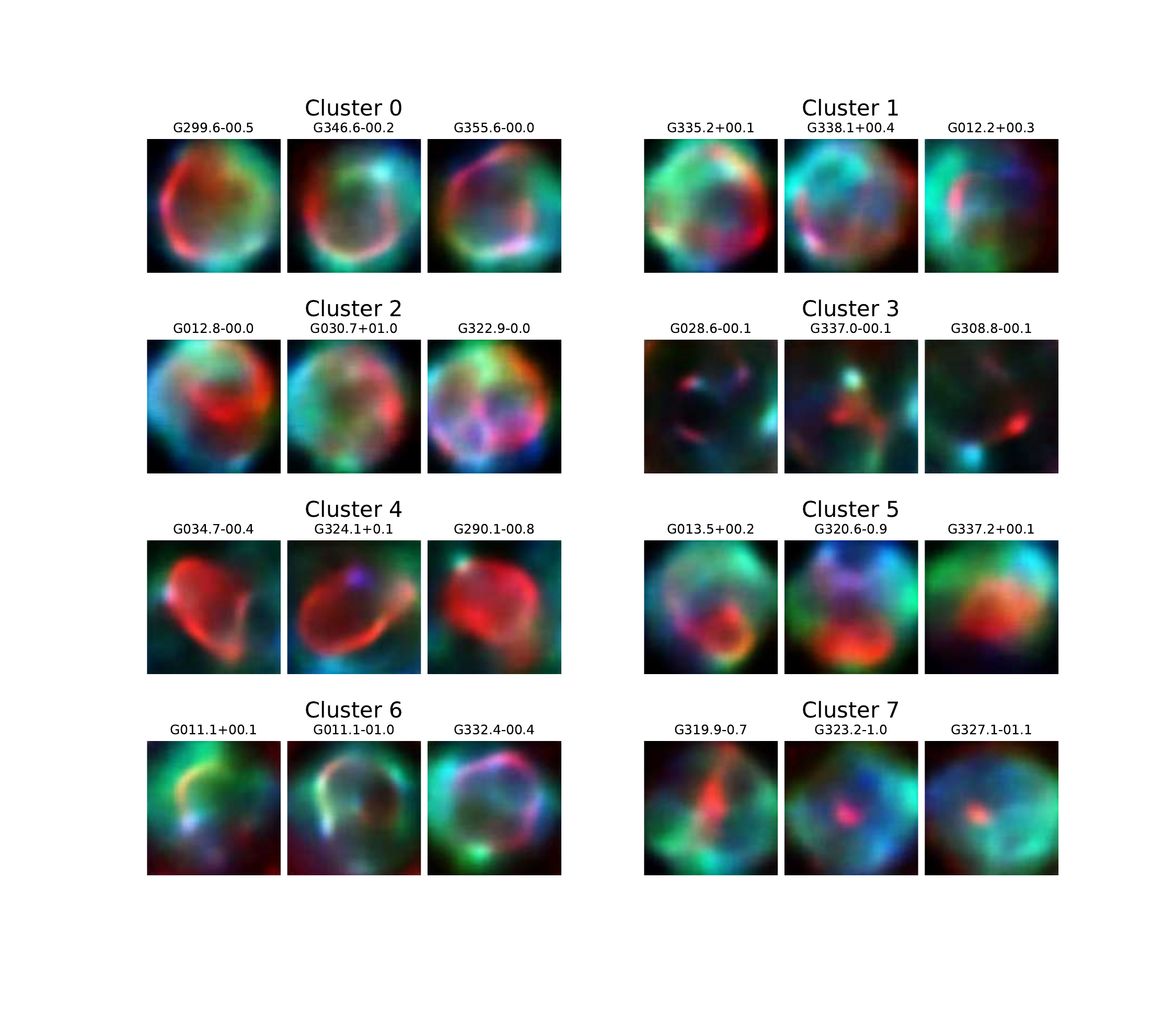}}
    \caption{Clustering results obtained with experiment B3 in Table \ref{tab:experiments}. Panel(a): Visualisation of DBSCAN clustering of experiment B3 SNR embeddings in a 6D manifold. A further 2D UMAP projection has been used for representation purposes only, with min\_dist=0.1 and n\_neighbors=15. Objects with label $-1$ are outliers. Panel(b): For each cluster found by DBSCAN, three representative SNRs were chosen and are presented as model reconstructions to show their main features.}
    \label{fig:best-cluster}
\end{figure*}
As mentioned above, clusterings are stable, which means that the elements belonging to the clusters are the same. We therefore picked experiment B3 as a reference for a discussion of their possible physical meaning.
Figure \ref{fig:best-cluster}(a) shows experiment B3, which corresponds to  a 6D UMAP projection and $\epsilon=0.40$. 
Eight  clusters can be identified, a representative of which is shown in Fig. \ref{fig:best-cluster}(b). 
At a glance, it is clear that both the morphology and colour distribution have played a role in defining the clusters. 
For some clusters, it is easier to spot a series of common characteristics: for example, cluster 4 is mostly made of SNRs where radio emission is predominant, with a limb-brightened elongated morphology that stands out over a more `uniform' IR emission; whilst  sources in cluster 5 display a more compact and irregular central radio emission. 
In the other clusters, the contribution from IR is stronger in general: cluster 6 contains sources with a rather complex morphology,  with filamentary IR emission that is roughly co-spatial with the radio; clusters 0 and 1 contain sources with a more diffuse IR component, and  shell-shaped or arc-like  radio emission. 
Cluster 3 displays objects with very localised strong emission both in IR and radio band. 
Cluster 7, on the other hand, is dominated by sources with a central and compact radio emission with no counterpart in the IR. 
Finally, cluster 2 is the most heterogeneous, showing no obvious common patterns.

\begin{figure*}
    \centering
    \includegraphics[width=0.9\textwidth]{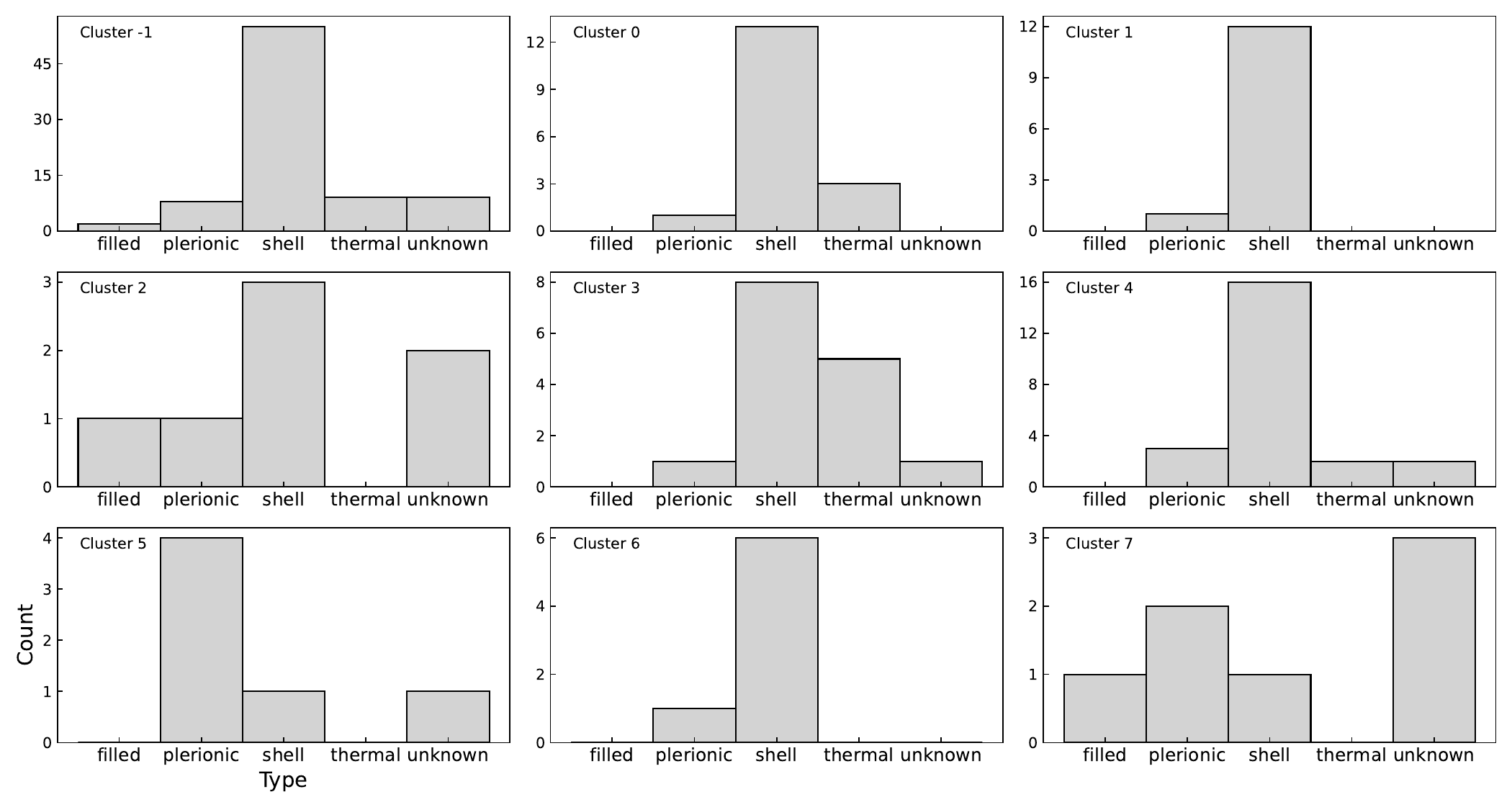}
    \caption{Clusters histograms. The distribution of SNRs in each cluster using the {\it classical} classification.}
    \label{fig:clust_histo}
\end{figure*}

\begin{figure*}
    \centering
    \includegraphics[width=0.9\textwidth]{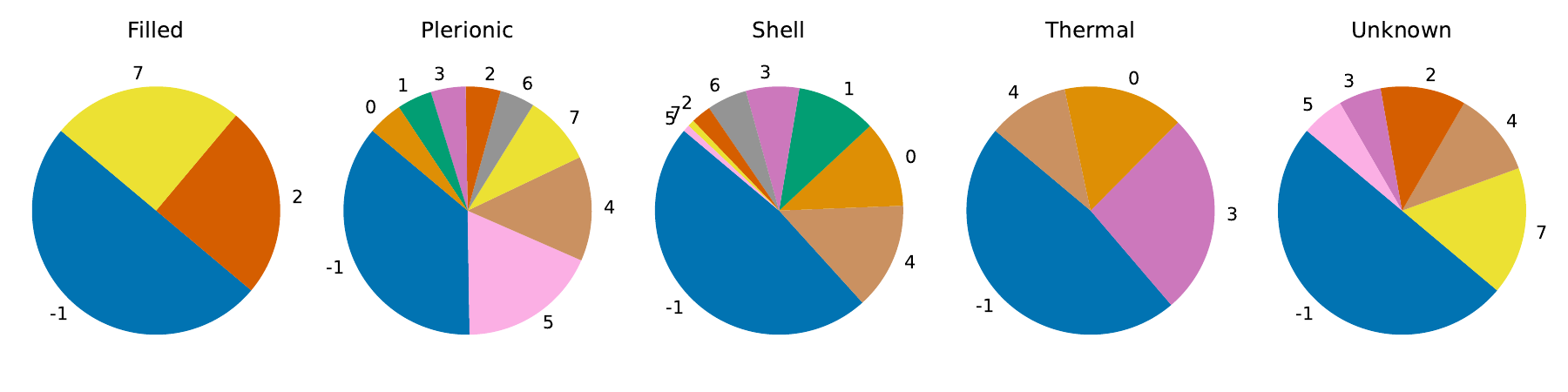}
    \caption{Distributions of SNRs with identical {\it classical} classification among different clusters (the colour palette for clusters is as in Fig.\ref{fig:best-cluster}).}
    \label{fig:clust_pie}
\end{figure*}

Providing a physical interpretation for the aforementioned clusters is not a straightforward task. 
As a starting point to investigate the physical processes responsible for the observed distribution and thus to characterise each cluster, we take advantage of the existing classification scheme in the SNR Manitoba catalogue.
In Fig. \ref{fig:clust_histo}, we show the distribution of SNR types in each cluster and in Fig. \ref{fig:clust_pie} we show the distribution of each type among the different clusters.
However, there is a factor that we must consider: such a classification scheme of SNRs relies on the available radio imagery, which is limited for many sources ---in terms of angular resolution and sensitivity--- compared to the new MeerKAT data. As a consequence, for instance, it is noticeable that shell-like SNRs are the most frequent component in all clusters, being the most numerous class in the catalogues.
In other words, the SMGPS images can reveal previously unseen structures and details ---such as regions of shallow radio emission--- that may affect the (morphology-based) classification of a given SNR (\citealt{Loru}).

In cluster 4, as mentioned, radio is the dominant emission and SNR members show a filled, limb-brightened morphology, with a weak but non-negligible diffuse central emission. 
However, the catalogue \citep{Ferrand2012} lists most of them  as shell-like sources, certainly due to the mere lack of sensitivity in previous observations. 
As seen from the example in Fig.\ref{fig:best-cluster}(b), their morphology resembles that of W44 (G034.7-00.4, \citealt{Dubner2017}), a well-studied composite SNR. Composite SNRs are objects with two different components emitting in radio for two simultaneous but different processes: in thermal composite or mixed-morphology (MM) SNRs, we observe a non-thermal emission from the expanding shell due to the interaction of the SN blast wave and the interstellar material (typically dense molecular clouds), and a central emission associated with a thermal X-ray detection; in plerionic composite or filled-centre SNRs, we again observe the synchrotron emission from the shell, plus a central non-thermal emission due to the presence of a pulsar wind nebula (PWN), that is, due to the wind of relativistic particles from the central spinning neutron star.\\
Although W44 is reported in the catalogue as plerionic composite, the very detailed study of \citet{CastellettiW44} revealed that there is no evidence in the radio continuum spectrum of any coupling between the associated pulsar and the SNR emission; that is, the central diffuse emission is not from the PWN, while thermal X-ray emission has been detected (as typically seen in MM SNRs).\\
Cluster 4 contains a further two thermal composite SNRs,  G290.1-00.8 and G327.4+00.4; while the nature of G008.7-00.1 and G338.3-00.0, reported as plerionic composite, is still debated, with authors claiming evidence of strong interaction with the surrounding molecular clouds (see \citealt{CastroW30} and \citealt{Lau}, respectively). 
It is therefore possible that a MM nature is the common denominator for this group, and that, having recovered a central, thus-far unseen emission  for most of them, the clustering is recognising this as a common feature.\\ 
Similarly, sources in the smaller cluster 7 have a more compact, slightly elongated, bright radio emission in the centre. Three out of four already classified SNRs are filled/plerionic composite SNRs (G029.7-00.3, G327.1-01.1 and G328.4+00.2), thus also in this case we could conclude that the central compact emission, and therefore the presence of a central PWN, could be common features of this cluster.

Cluster 6 is composed exclusively of shell-like SNRs. The only exception could be
G011+00.1, whose nature as composite plerionic is described as uncertain in the catalogue and whose radio morphology recovered from MeerKAT imagery does not show any strong compact or diffuse central emission.
Sources in cluster 6 have prominent dust features visible as shell-like structures or filaments at mid-infrared (MIR) and FIR wavelengths that are not always co-spatial with the radio. Four of these sources (G011.1+00.1, G011.1-01.0, G332.4-00.4, and G340.6+00.3) are listed as dusty SNRs by \cite{Chawner20}.

Cluster 3 contains sources that present bright features at IR and radio wavelengths. 
From Fig.\ref{fig:clust_pie}, we see that most of the thermal composite SNRs are included in this cluster, pointing to a possible thermal origin of both the emissions. Thermal composite SNRs are associated with gamma-ray sources mainly exhibiting a hadronic nature; that is, relativistic particles within these sources collide with dense ambient targets, producing pions that subsequently decay into gamma rays. Theoretical modelling suggests the presence of remnants expanding within a diffuse environment, with small ISM clouds swallowed up by the main shock front, and  heated up to X-ray-emitting temperatures. 
While we cannot firmly exclude that some of these features are related to artefacts caused by the subtraction of compact sources, these bright regions may be the effect of such interaction between the SNR and close molecular clouds.

Clusters 0 and 1 lay in two opposite positions on the visualisation plane (Fig.\ref{fig:best-cluster}-a), but their members show very similar characteristics: SNRs have a clear shell-like shape or arcs in the radio band, while IR emissions are more diffuse. However, the origin of the latter is more difficult to infer: IR emissions may be from the SNR itself, or they may be a contribution from the ISM. Distribution plots show that these clusters are mostly constituted from shell-like SNRs.

Finally, clusters 2 and 5 do not show any clear common pattern among their relative members, appearing heterogeneous at least to the human eye. 
Even from the classical classification plots (Figs. \ref{fig:clust_histo} and \ref{fig:clust_pie}), we can see that they are small clusters with a varied composition.

\section{Conclusions and future work}
\label{sec:conclusions}

In this work, we explore the use of unsupervised machine learning techniques to analyse the multi-wavelength properties of a representative sample of Galactic SNRs using high-resolution IR and radio images from the WISE, Hi-GAL, and SMGPS surveys. We fed the images to a CAE to produce a compact representation, and then searched the latent space for physically meaningful clusters using a density-based clustering approach. The key findings of this work are summarised as follows:
\begin{itemize}
    \item Despite the reduced number of example sources in the analysed sample ($\sim180$ SNRs), the CAE produces proper reconstructions that capture the most important features of the input images. The presence of masked regions in the input images (corresponding to compact sources not related to the SNRs) may influence the latent space structure, even though the CAE interpolates them in the learning process, providing continuity with the neighbouring pixels. The overall impact of masked sources is difficult to calibrate (see Appendix B), but we acknowledge that some corner cases (e.g. very bright compact sources that deviate from a Gaussian profile for which masking may not be perfect) tend to produce artefacts around the masked regions.
    \item The resulting latent space has a sparse topology, and therefore the performance of density-based clustering algorithms is poor ---they are generally unable to find meaningful clusters.
    \item The projection of the autoencoded embeddings into a lower dimensionality space greatly improves the performance of the clustering stage. In particular, following the so-called N2D approach, UMAP can effectively separate SNRs into distinct groups based on their multi-wavelength properties. We tested a wide range of possible manifold dimensions (from 4 to 12), ideally correlated with the expected number of clusters. This approach results in a more structured  ---but still sparse--- feature space, in which DBSCAN is able to isolate meaningful clusters. 
    \item For a representative clustering solution in an eight-dimensional manifold, we find eight well-defined clusters, while almost half of the objects remain as outliers. The analysis of the general properties of the clusters provides hints as to their connection to physical features of the SNRs and their close surroundings, such as a relation to the morphological SNR type, the presence of PWNe within the remnants, the existence of prominent dust features correlated with the radio (e.g. filaments or shells), and the possible interaction with neighbouring molecular clouds.
\end{itemize}

While the methods employed here have been widely used to address other astrophysical problems, mostly constrained to extragalactic science, this work represents a first, pioneering step in the application of an image feature extraction$+$clustering pipeline to well-resolved Galactic sources. As such, it is necessary to deal with a series of problems intrinsic to Galactic science, such as a limited number of samples and the contamination or confusion due to Galactic plane diffuse emission. Unfortunately, these two issues are entangled: the number of known SNRs and other extended radio sources is expected to grow substantially in the coming years thanks to the higher sensitivity of MeerKAT and other Square Kilometre Array (SKA) precursors (e.g. see \citealt{Bordiu}), but at the cost of detecting shallower diffuse emission that increases confusion at frequencies around $\sim$1 GHz. We note that, regardless, the architecture presented here is highly adaptable and can therefore be applied to study other Galactic sources (e.g. H\textsc{ii} regions, unclassified sources).

There is still a large fraction of sources for which no cluster can be reliably assigned with the available data. Somehow, this points out the limitations of the methods employed, as density-based clustering algorithms are unable to find better partitions of the data. 
However, we stress again that the intrinsic structure of the feature space is sparse. 
In other words, there is a `continuum' of features rather than a clear distinction between groups. 
As a consequence, many of the unclassified objects likely share the properties of several adjacent clusters. It is possible that the addition of complementary information coming from different wavelength ranges (e.g. the X-ray data to be delivered by eROSITA; \citealt{2012arXiv1209.3114M}) or physical parameter distribution maps (e.g. dust temperature or optical depth distribution maps) will mitigate the number of outliers by introducing new key features related to the most energetic processes ---that are indeed relevant to the study of SNRs. Such new features will also facilitate the physical interpretation of the clusters.

Another crucial aspect that needs to be addressed in the future is data normalisation. The `min-max' per channel normalisation used in this work removes valuable physical information, hampering the use of the channel intensity ratios as useful diagnostic tools for measuring the correlation between warm dust and hot plasma. 

In any case, the lack of zero-baseline observations in SMGPS data presents a challenge for any analysis based on ratios between pixel values. Negative pixels around bright sources, which are unphysical, already render channel-intensity ratios involving radio maps useless. Unfortunately, to the best of our knowledge, there are no plans to mitigate this effect in future SMGPS data releases, either by complementing the interferometric data with single-dish observations or by applying a data reduction strategy more suited for extended structures. Some SKA precursor projects, such as ASKAP's Evolutionary Map of the Universe \citep{EMU} performed with the Australian Square Kilometre Array Pathfinder, are in the process of adding complementary zero-baseline observations to their interferometric maps (e.g. PEGASUS project). However, it could be worth investigating alternative methods for handling negative pixel values in a way that allows the retention of physical information.

While all the aforementioned issues clearly limit our ability to provide an in-depth analysis of Galactic SNRs from a multi-wavelength unsupervised perspective, the results obtained so far are nonetheless promising, strongly suggesting the existence of unseen `patterns' in the population that go beyond their mere morphological features in the radio band. In a future work, we plan to extend the presented methodology, incorporating new data and testing new deep architectures with the aim of obtaining a more robust clusterisation.

\section*{Data availability}

The pipeline presented in this work, initially built \textit{ad hoc} for this science case, has been revamped in the context of the H2020 NEANIAS project, and it is now available as a general-purpose cloud service, named \textsc{latent space explorer}\footnote{https://lse.neanias.eu/}, that includes a visualization module for exploratory analysis of the resulting clusters (see \citealt{Cecconello22} for further details).

\begin{acknowledgements}
F.B and C.B. acknowledge support from the European Commission Horizon 2020 research and innovation programme under the grant agreement No. 863448 (NEANIAS).\\
Supported by Italian Research Center on High Performance Computing Big Data and Quantum Computing (ICSC), project funded by European Union - NextGenerationEU - and National Recovery and Resilience Plan (NRRP) - Mission 4 Component 2 within the activities of Spoke 3 (Astrophysics and Cosmos Observations).\\
The MeerKAT telescope is operated by the South African Radio Astronomy Observatory, which is a facility of the National Research Foundation, an agency of the Department of Science and Innovation.
The National Radio Astronomy Observatory is a facility of the National Science Foundation operated under cooperative agreement by Associated Universities, Inc. The Centre for Astrophysics Research at the University of Hertfordshire kindly provided access to their HPC facilities for data processing and storage.\\
This research made use of Montage. It is funded by the National Science Foundation under Grant Number ACI-1440620, and was previously funded by the National Aeronautics and Space Administration's Earth Science Technology Office, Computation Technologies Project, under Cooperative Agreement Number NCC5-626 between NASA and the California Institute of Technology. 
\end{acknowledgements}

%-------------------------------------------------------------------
\bibliographystyle{aa}
\bibliography{libreria}

\begin{thebibliography}{49}
\expandafter\ifx\csname natexlab\endcsname\relax\def\natexlab#1{#1}\fi

\bibitem[{{Ball} \& {Brunner}(2010)}]{2010IJMPD..19.1049B}
{Ball}, N.~M. \& {Brunner}, R.~J. 2010, International Journal of Modern Physics D, 19, 1049

\bibitem[{{Bloom} {et~al.}(2012){Bloom}, {Richards}, {Nugent}, {Quimby}, {Kasliwal}, {Starr}, {Poznanski}, {Ofek}, {Cenko}, {Butler}, {Kulkarni}, {Gal-Yam}, \& {Law}}]{Bloom2012}
{Bloom}, J.~S., {Richards}, J.~W., {Nugent}, P.~E., {et~al.} 2012, \pasp, 124, 1175

\bibitem[{{Bordiu} {et~al.}(2024){Bordiu}, {Riggi}, \& {Bufano}}]{Bordiu}
{Bordiu}, C., {Riggi}, S., \& {Bufano}, F. 2024, \aap, submitted

\bibitem[{{Brescia} {et~al.}(2021){Brescia}, {Cavuoti}, {Razim}, {Amaro}, {Riccio}, \& {Longo}}]{Brescia2021}
{Brescia}, M., {Cavuoti}, S., {Razim}, O., {et~al.} 2021, Frontiers in Astronomy and Space Sciences, 8, 70

\bibitem[{Caliński \& Harabasz(1974)}]{CalinskiHarabasz1974}
Caliński, T. \& Harabasz, J. 1974, Communications in Statistics, 3, 1

\bibitem[{{Castelletti} {et~al.}(2007){Castelletti}, {Dubner}, {Brogan}, \& {Kassim}}]{CastellettiW44}
{Castelletti}, G., {Dubner}, G., {Brogan}, C., \& {Kassim}, N.~E. 2007, \aap, 471, 537

\bibitem[{{Castro} \& {Slane}(2010)}]{CastroW30}
{Castro}, D. \& {Slane}, P. 2010, \apj, 717, 372

\bibitem[{{Cecconello} {et~al.}(2022){Cecconello}, {Bordiu}, {Bufano}, {Puerari}, {Riggi}, {Schisano}, {Sciacca}, {Maruccia}, \& {Vizzari}}]{Cecconello22}
{Cecconello}, T., {Bordiu}, C., {Bufano}, F., {et~al.} 2022, arXiv e-prints, arXiv:2204.13933

\bibitem[{{Chawner} {et~al.}(2020){Chawner}, {Gomez}, {Matsuura}, {Smith}, {Papageorgiou}, {Rho}, {Noriega-Crespo}, {De Looze}, {Barlow}, {Cigan}, {Dunne}, \& {Marsh}}]{Chawner20}
{Chawner}, H., {Gomez}, H.~L., {Matsuura}, M., {et~al.} 2020, \mnras, 493, 2706

\bibitem[{{Cheng} {et~al.}(2020){Cheng}, {Li}, {Conselice}, {Arag{\'o}n-Salamanca}, {Dye}, \& {Metcalf}}]{Cheng2020}
{Cheng}, T.-Y., {Li}, N., {Conselice}, C.~J., {et~al.} 2020, \mnras, 494, 3750

\bibitem[{Davies \& Bouldin(1979)}]{DaviesBoulding79}
Davies, D.~L. \& Bouldin, D.~W. 1979, IEEE Transactions on Pattern Analysis and Machine Intelligence, PAMI-1, 224

\bibitem[{{Dieleman} {et~al.}(2015){Dieleman}, {Willett}, \& {Dambre}}]{Dieleman2015}
{Dieleman}, S., {Willett}, K.~W., \& {Dambre}, J. 2015, \mnras, 450, 1441

\bibitem[{{Dubner}(2017)}]{Dubner2017}
{Dubner}, G. 2017, in Handbook of Supernovae, ed. A.~W. {Alsabti} \& P.~{Murdin}, 2041

\bibitem[{Ester {et~al.}(1996)Ester, Kriegel, Sander, \& Xu}]{ester1996densitybased}
Ester, M., Kriegel, H.-P., Sander, J., \& Xu, X. 1996, in Proc. of 2nd International Conference on Knowledge Discovery and, 226--231

\bibitem[{{Ferrand} \& {Safi-Harb}(2012)}]{Ferrand2012}
{Ferrand}, G. \& {Safi-Harb}, S. 2012, Advances in Space Research, 49, 1313

\bibitem[{{Fraix-Burnet} {et~al.}(2021){Fraix-Burnet}, {Bouveyron}, \& {Moultaka}}]{Fraix2021}
{Fraix-Burnet}, D., {Bouveyron}, C., \& {Moultaka}, J. 2021, \aap, 649, A53

\bibitem[{{Goedhart} {et~al.}(2024){Goedhart}, {Cotton}, {Camilo}, {Thompson}, {Umana}, {Bietenholz}, {Woudt}, {Anderson}, {Bordiu}, {Buckley}, {Buemi}, {Bufano}, {Cavallaro}, {Chen}, {Chibueze}, {Egbo}, {Frank}, {Hoare}, {Ingallinera}, {Irabor}, {Kraan-Korteweg}, {Kurapati}, {Leto}, {Loru}, {Mutale}, {Obonyo}, {Plavin}, {Rajohnson}, {Rigby}, {Riggi}, {Seidu}, {Serra}, {Smart}, {Stappers}, {Steyn}, {Surnis}, {Trigilio}, {Williams}, {Abbott}, {Adam}, {Asad}, {Baloyi}, {Bauermeister}, {Bennet}, {Bester}, {Botha}, {Brederode}, {Buchner}, {Burger}, {Cheetham}, {Cloete}, {de Villiers}, {de Villiers}, {du Toit}, {Esterhuyse}, {Fanaroff}, {Fourie}, {Gamatham}, {Gatsi}, {Geyer}, {Gouws}, {Gumede}, {Heywood}, {Hokwana}, {Hoosen}, {Horn}, {Horrell}, {Hugo}, {Isaacson}, {J{\'o}zsa}, {Jonas}, {Jordaan}, {Joubert}, {Julie}, {Kapp}, {Kriek}, {Kriel}, {Krishnan}, {Kusel}, {Legodi}, {Lehmensiek}, {Lord}, {Macfarlane}, {Magnus}, {Magozore}, {Main}, {Malan}, {Manley}, {Marais}, {Maree}, {Martens}, {Maruping}, {McAlpine},
  {Merry}, {Mgodeli}, {Millenaar}, {Mokone}, {Monama}, {New}, {Ngcebetsha}, {Ngoasheng}, {Nicolson}, {Ockards}, {Oozeer}, {Passmoor}, {Patel}, {Peens-Hough}, {Perkins}, {Ramaila}, {Ratcliffe}, {Renil}, {Richter}, {Salie}, {Sambu}, {Schollar}, {Schwardt}, {Schwartz}, {Serylak}, {Siebrits}, {Sirothia}, {Slabber}, {Smirnov}, {Tiplady}, {van Balla}, {van der Byl}, {Van Tonder}, {Venter}, {Venter}, {Welz}, \& {Williams}}]{Goedhart}
{Goedhart}, S., {Cotton}, W.~D., {Camilo}, F., {et~al.} 2024, \mnras, 531, 649

\bibitem[{{Goldstein} {et~al.}(2015){Goldstein}, {D'Andrea}, {Fischer}, {Foley}, {Gupta}, {Kessler}, {Kim}, {Nichol}, {Nugent}, {Papadopoulos}, {Sako}, {Smith}, {Sullivan}, {Thomas}, {Wester}, {Wolf}, {Abdalla}, {Banerji}, {Benoit-L{\'e}vy}, {Bertin}, {Brooks}, {Carnero Rosell}, {Castander}, {da Costa}, {Covarrubias}, {DePoy}, {Desai}, {Diehl}, {Doel}, {Eifler}, {Fausti Neto}, {Finley}, {Flaugher}, {Fosalba}, {Frieman}, {Gerdes}, {Gruen}, {Gruendl}, {James}, {Kuehn}, {Kuropatkin}, {Lahav}, {Li}, {Maia}, {Makler}, {March}, {Marshall}, {Martini}, {Merritt}, {Miquel}, {Nord}, {Ogando}, {Plazas}, {Romer}, {Roodman}, {Sanchez}, {Scarpine}, {Schubnell}, {Sevilla-Noarbe}, {Smith}, {Soares-Santos}, {Sobreira}, {Suchyta}, {Swanson}, {Tarle}, {Thaler}, \& {Walker}}]{Goldstein2015}
{Goldstein}, D.~A., {D'Andrea}, C.~B., {Fischer}, J.~A., {et~al.} 2015, \aj, 150, 82

\bibitem[{{Green}(2019)}]{Green2019}
{Green}, D.~A. 2019, Journal of Astrophysics and Astronomy, 40, 36

\bibitem[{{Griffin} {et~al.}(2010){Griffin}, {Abergel}, {Abreu}, {Ade}, {Andr{\'e}}, {Augueres}, {Babbedge}, {Bae}, {Baillie}, {Baluteau}, {Barlow}, {Bendo}, {Benielli}, {Bock}, {Bonhomme}, {Brisbin}, {Brockley-Blatt}, {Caldwell}, {Cara}, {Castro-Rodriguez}, {Cerulli}, {Chanial}, {Chen}, {Clark}, {Clements}, {Clerc}, {Coker}, {Communal}, {Conversi}, {Cox}, {Crumb}, {Cunningham}, {Daly}, {Davis}, {de Antoni}, {Delderfield}, {Devin}, {di Giorgio}, {Didschuns}, {Dohlen}, {Donati}, {Dowell}, {Dowell}, {Duband}, {Dumaye}, {Emery}, {Ferlet}, {Ferrand}, {Fontignie}, {Fox}, {Franceschini}, {Frerking}, {Fulton}, {Garcia}, {Gastaud}, {Gear}, {Glenn}, {Goizel}, {Griffin}, {Grundy}, {Guest}, {Guillemet}, {Hargrave}, {Harwit}, {Hastings}, {Hatziminaoglou}, {Herman}, {Hinde}, {Hristov}, {Huang}, {Imhof}, {Isaak}, {Israelsson}, {Ivison}, {Jennings}, {Kiernan}, {King}, {Lange}, {Latter}, {Laurent}, {Laurent}, {Leeks}, {Lellouch}, {Levenson}, {Li}, {Li}, {Lilienthal}, {Lim}, {Liu}, {Lu}, {Madden}, {Mainetti}, {Marliani},
  {McKay}, {Mercier}, {Molinari}, {Morris}, {Moseley}, {Mulder}, {Mur}, {Naylor}, {Nguyen}, {O'Halloran}, {Oliver}, {Olofsson}, {Olofsson}, {Orfei}, {Page}, {Pain}, {Panuzzo}, {Papageorgiou}, {Parks}, {Parr-Burman}, {Pearce}, {Pearson}, {P{\'e}rez-Fournon}, {Pinsard}, {Pisano}, {Podosek}, {Pohlen}, {Polehampton}, {Pouliquen}, {Rigopoulou}, {Rizzo}, {Roseboom}, {Roussel}, {Rowan-Robinson}, {Rownd}, {Saraceno}, {Sauvage}, {Savage}, {Savini}, {Sawyer}, {Scharmberg}, {Schmitt}, {Schneider}, {Schulz}, {Schwartz}, {Shafer}, {Shupe}, {Sibthorpe}, {Sidher}, {Smith}, {Smith}, {Smith}, {Spencer}, {Stobie}, {Sudiwala}, {Sukhatme}, {Surace}, {Stevens}, {Swinyard}, {Trichas}, {Tourette}, {Triou}, {Tseng}, {Tucker}, {Turner}, {Vaccari}, {Valtchanov}, {Vigroux}, {Virique}, {Voellmer}, {Walker}, {Ward}, {Waskett}, {Weilert}, {Wesson}, {White}, {Whitehouse}, {Wilson}, {Winter}, {Woodcraft}, {Wright}, {Xu}, {Zavagno}, {Zemcov}, {Zhang}, \& {Zonca}}]{SPIRE}
{Griffin}, M.~J., {Abergel}, A., {Abreu}, A., {et~al.} 2010, \aap, 518, L3

\bibitem[{{Iwasaki} {et~al.}(2019){Iwasaki}, {Ichinohe}, \& {Uchiyama}}]{Iwasaki2019}
{Iwasaki}, H., {Ichinohe}, Y., \& {Uchiyama}, Y. 2019, \mnras, 488, 4106

\bibitem[{{Jacobs} {et~al.}(2017){Jacobs}, {Glazebrook}, {Collett}, {More}, \& {McCarthy}}]{Jacobs2017}
{Jacobs}, C., {Glazebrook}, K., {Collett}, T., {More}, A., \& {McCarthy}, C. 2017, \mnras, 471, 167

\bibitem[{Khachiyan(1980)}]{khachiyan}
Khachiyan, L.~G. 1980, USSR Computational Mathematics and Mathematical Physics, 20, 53

\bibitem[{{Lanusse} {et~al.}(2018){Lanusse}, {Ma}, {Li}, {Collett}, {Li}, {Ravanbakhsh}, {Mandelbaum}, \& {P{\'o}czos}}]{Lanusse2018}
{Lanusse}, F., {Ma}, Q., {Li}, N., {et~al.} 2018, \mnras, 473, 3895

\bibitem[{{Lau} {et~al.}(2017){Lau}, {Rowell}, {Burton}, {Fukui}, {Aharonian}, {Oya}, {Vink}, {Ohm}, \& {Casanova}}]{Lau}
{Lau}, J.~C., {Rowell}, G., {Burton}, M.~G., {et~al.} 2017, \mnras, 464, 3757

\bibitem[{{Lopez} {et~al.}(2009){Lopez}, {Ramirez-Ruiz}, {Badenes}, {Huppenkothen}, {Jeltema}, \& {Pooley}}]{Lopez2009}
{Lopez}, L.~A., {Ramirez-Ruiz}, E., {Badenes}, C., {et~al.} 2009, \apjl, 706, L106

\bibitem[{{Loru} {et~al.}(2024){Loru}, {Ingallinera}, \& {Umana}}]{Loru}
{Loru}, S., {Ingallinera}, A., \& {Umana}, G. 2024, \aap, submitted

\bibitem[{McConville {et~al.}(2020)McConville, Santos-Rodriguez, Piechocki, \& Craddock}]{McConville2020}
McConville, R., Santos-Rodriguez, R., Piechocki, R.~J., \& Craddock, I. 2020, in 25th International Conference on Pattern Recognition, {ICPR} 2020 ({IEEE} Computer Society)

\bibitem[{McInnes {et~al.}(2020)McInnes, Healy, \& Melville}]{umap}
McInnes, L., Healy, J., \& Melville, J. 2020, UMAP: Uniform Manifold Approximation and Projection for Dimension Reduction

\bibitem[{{Merloni} {et~al.}(2012){Merloni}, {Predehl}, {Becker}, {B{\"o}hringer}, {Boller}, {Brunner}, {Brusa}, {Dennerl}, {Freyberg}, {Friedrich}, {Georgakakis}, {Haberl}, {Hasinger}, {Meidinger}, {Mohr}, {Nandra}, {Rau}, {Reiprich}, {Robrade}, {Salvato}, {Santangelo}, {Sasaki}, {Schwope}, {Wilms}, \& {German eROSITA Consortium}}]{2012arXiv1209.3114M}
{Merloni}, A., {Predehl}, P., {Becker}, W., {et~al.} 2012, arXiv e-prints, arXiv:1209.3114

\bibitem[{{Molinari} {et~al.}(2010){Molinari}, {Swinyard}, {Bally}, {Barlow}, {Bernard}, {Martin}, {Moore}, {Noriega-Crespo}, {Plume}, {Testi}, {Zavagno}, {Abergel}, {Ali}, {Anderson}, {Andr{\'e}}, {Baluteau}, {Battersby}, {Beltr{\'a}n}, {Benedettini}, {Billot}, {Blommaert}, {Bontemps}, {Boulanger}, {Brand}, {Brunt}, {Burton}, {Calzoletti}, {Carey}, {Caselli}, {Cesaroni}, {Cernicharo}, {Chakrabarti}, {Chrysostomou}, {Cohen}, {Compiegne}, {de Bernardis}, {de Gasperis}, {di Giorgio}, {Elia}, {Faustini}, {Flagey}, {Fukui}, {Fuller}, {Ganga}, {Garcia-Lario}, {Glenn}, {Goldsmith}, {Griffin}, {Hoare}, {Huang}, {Ikhenaode}, {Joblin}, {Joncas}, {Juvela}, {Kirk}, {Lagache}, {Li}, {Lim}, {Lord}, {Marengo}, {Marshall}, {Masi}, {Massi}, {Matsuura}, {Minier}, {Miville-Desch{\^e}nes}, {Montier}, {Morgan}, {Motte}, {Mottram}, {M{\"u}ller}, {Natoli}, {Neves}, {Olmi}, {Paladini}, {Paradis}, {Parsons}, {Peretto}, {Pestalozzi}, {Pezzuto}, {Piacentini}, {Piazzo}, {Polychroni}, {Pomar{\`e}s}, {Popescu}, {Reach}, {Ristorcelli},
  {Robitaille}, {Robitaille}, {Rod{\'o}n}, {Roy}, {Royer}, {Russeil}, {Saraceno}, {Sauvage}, {Schilke}, {Schisano}, {Schneider}, {Schuller}, {Schulz}, {Sibthorpe}, {Smith}, {Smith}, {Spinoglio}, {Stamatellos}, {Strafella}, {Stringfellow}, {Sturm}, {Taylor}, {Thompson}, {Traficante}, {Tuffs}, {Umana}, {Valenziano}, {Vavrek}, {Veneziani}, {Viti}, {Waelkens}, {Ward-Thompson}, {White}, {Wilcock}, {Wyrowski}, {Yorke}, \& {Zhang}}]{HIGAL}
{Molinari}, S., {Swinyard}, B., {Bally}, J., {et~al.} 2010, \aap, 518, L100

\bibitem[{{Naul} {et~al.}(2018){Naul}, {Bloom}, {P{\'e}rez}, \& {van der Walt}}]{Naul2018}
{Naul}, B., {Bloom}, J.~S., {P{\'e}rez}, F., \& {van der Walt}, S. 2018, Nature Astronomy, 2, 151

\bibitem[{{Norris} {et~al.}(2011){Norris}, {Hopkins}, {Afonso}, {Brown}, {Condon}, {Dunne}, {Feain}, {Hollow}, {Jarvis}, {Johnston-Hollitt}, {Lenc}, {Middelberg}, {Padovani}, {Prandoni}, {Rudnick}, {Seymour}, {Umana}, {Andernach}, {Alexander}, {Appleton}, {Bacon}, {Banfield}, {Becker}, {Brown}, {Ciliegi}, {Jackson}, {Eales}, {Edge}, {Gaensler}, {Giovannini}, {Hales}, {Hancock}, {Huynh}, {Ibar}, {Ivison}, {Kennicutt}, {Kimball}, {Koekemoer}, {Koribalski}, {L{\'o}pez-S{\'a}nchez}, {Mao}, {Murphy}, {Messias}, {Pimbblet}, {Raccanelli}, {Randall}, {Reiprich}, {Roseboom}, {R{\"o}ttgering}, {Saikia}, {Sharp}, {Slee}, {Smail}, {Thompson}, {Urquhart}, {Wall}, \& {Zhao}}]{EMU}
{Norris}, R.~P., {Hopkins}, A.~M., {Afonso}, J., {et~al.} 2011, \pasa, 28, 215

\bibitem[{{Pilbratt} {et~al.}(2010){Pilbratt}, {Riedinger}, {Passvogel}, {Crone}, {Doyle}, {Gageur}, {Heras}, {Jewell}, {Metcalfe}, {Ott}, \& {Schmidt}}]{pilbratt}
{Pilbratt}, G.~L., {Riedinger}, J.~R., {Passvogel}, T., {et~al.} 2010, \aap, 518, L1

\bibitem[{{Pinheiro Gon{\c{c}}alves} {et~al.}(2011){Pinheiro Gon{\c{c}}alves}, {Noriega-Crespo}, {Paladini}, {Martin}, \& {Carey}}]{Pinheiro}
{Pinheiro Gon{\c{c}}alves}, D., {Noriega-Crespo}, A., {Paladini}, R., {Martin}, P.~G., \& {Carey}, S.~J. 2011, \aj, 142, 47

\bibitem[{{Poglitsch} {et~al.}(2010){Poglitsch}, {Waelkens}, {Geis}, {Feuchtgruber}, {Vandenbussche}, {Rodriguez}, {Krause}, {Renotte}, {van Hoof}, {Saraceno}, {Cepa}, {Kerschbaum}, {Agn{\`e}se}, {Ali}, {Altieri}, {Andreani}, {Augueres}, {Balog}, {Barl}, {Bauer}, {Belbachir}, {Benedettini}, {Billot}, {Boulade}, {Bischof}, {Blommaert}, {Callut}, {Cara}, {Cerulli}, {Cesarsky}, {Contursi}, {Creten}, {De Meester}, {Doublier}, {Doumayrou}, {Duband}, {Exter}, {Genzel}, {Gillis}, {Gr{\"o}zinger}, {Henning}, {Herreros}, {Huygen}, {Inguscio}, {Jakob}, {Jamar}, {Jean}, {de Jong}, {Katterloher}, {Kiss}, {Klaas}, {Lemke}, {Lutz}, {Madden}, {Marquet}, {Martignac}, {Mazy}, {Merken}, {Montfort}, {Morbidelli}, {M{\"u}ller}, {Nielbock}, {Okumura}, {Orfei}, {Ottensamer}, {Pezzuto}, {Popesso}, {Putzeys}, {Regibo}, {Reveret}, {Royer}, {Sauvage}, {Schreiber}, {Stegmaier}, {Schmitt}, {Schubert}, {Sturm}, {Thiel}, {Tofani}, {Vavrek}, {Wetzstein}, {Wieprecht}, \& {Wiezorrek}}]{PACS}
{Poglitsch}, A., {Waelkens}, C., {Geis}, N., {et~al.} 2010, \aap, 518, L2

\bibitem[{Reddi {et~al.}(2018)Reddi, Kale, \& Kumar}]{j.2018on}
Reddi, S.~J., Kale, S., \& Kumar, S. 2018, in International Conference on Learning Representations

\bibitem[{{Riggi} {et~al.}(2021{\natexlab{a}}){Riggi}, {Bordiu}, {Vitello}, {Tudisco}, {Sciacca}, {Magro}, {Sortino}, {Pino}, {Molinaro}, {Benedettini}, {Leurini}, {Bufano}, {Raciti}, \& {Becciani}}]{RiggiBordiu2021}
{Riggi}, S., {Bordiu}, C., {Vitello}, F., {et~al.} 2021{\natexlab{a}}, Astronomy and Computing, 37, 100506

\bibitem[{{Riggi} {et~al.}(2016){Riggi}, {Ingallinera}, {Leto}, {Cavallaro}, {Bufano}, {Schillir{\`o}}, {Trigilio}, {Umana}, {Buemi}, \& {Norris}}]{Riggi2016}
{Riggi}, S., {Ingallinera}, A., {Leto}, P., {et~al.} 2016, \mnras, 460, 1486

\bibitem[{{Riggi} {et~al.}(2021{\natexlab{b}}){Riggi}, {Umana}, {Trigilio}, {Cavallaro}, {Ingallinera}, {Leto}, {Bufano}, {Norris}, {Hopkins}, {Filipovi{\'c}}, {Andernach}, {van Loon}, {Micha{\l}owski}, {Bordiu}, {An}, {Buemi}, {Carretti}, {Collier}, {Joseph}, {Koribalski}, {Kothes}, {Loru}, {McConnell}, {Pommier}, {Sciacca}, {Schillir{\`o}}, {Vitello}, {Warhurst}, \& {Whiting}}]{RiggiUmana2021}
{Riggi}, S., {Umana}, G., {Trigilio}, C., {et~al.} 2021{\natexlab{b}}, \mnras, 502, 60

\bibitem[{{Rubin} \& {Gal-Yam}(2016)}]{Rubin2016}
{Rubin}, A. \& {Gal-Yam}, A. 2016, \apj, 828, 111

\bibitem[{{S{\'a}nchez Almeida} {et~al.}(2010){S{\'a}nchez Almeida}, {Aguerri}, {Mu{\~n}oz-Tu{\~n}{\'o}n}, \& {de Vicente}}]{SanchezAlmeida2010}
{S{\'a}nchez Almeida}, J., {Aguerri}, J.~A.~L., {Mu{\~n}oz-Tu{\~n}{\'o}n}, C., \& {de Vicente}, A. 2010, \apj, 714, 487

\bibitem[{{S{\'a}nchez Almeida} \& {Allende Prieto}(2013)}]{SanchezAlmeida2013}
{S{\'a}nchez Almeida}, J. \& {Allende Prieto}, C. 2013, \apj, 763, 50

\bibitem[{{Shallue} \& {Vanderburg}(2018)}]{Shallue2018}
{Shallue}, C.~J. \& {Vanderburg}, A. 2018, \aj, 155, 94

\bibitem[{{Sortino} {et~al.}(2023){Sortino}, {Magro}, {Fiameni}, {Sciacca}, {Riggi}, {DeMarco}, {Spampinato}, {Hopkins}, {Bufano}, {Schillir{\`o}}, {Bordiu}, \& {Pino}}]{Sortino}
{Sortino}, R., {Magro}, D., {Fiameni}, G., {et~al.} 2023, Experimental Astronomy, 56, 293

\bibitem[{{Spindler} {et~al.}(2021){Spindler}, {Geach}, \& {Smith}}]{Spindler2021}
{Spindler}, A., {Geach}, J.~E., \& {Smith}, M.~J. 2021, \mnras, 502, 985

\bibitem[{{Valenzuela} \& {Pichara}(2018)}]{Valenzuela2018}
{Valenzuela}, L. \& {Pichara}, K. 2018, \mnras, 474, 3259

\bibitem[{{Var{\'o}n} {et~al.}(2011){Var{\'o}n}, {Alzate}, {Suykens}, \& {Debosscher}}]{Varon2011}
{Var{\'o}n}, C., {Alzate}, C., {Suykens}, J.~A.~K., \& {Debosscher}, J. 2011, \aap, 531, A156

\bibitem[{{Wright} {et~al.}(2010){Wright}, {Eisenhardt}, {Mainzer}, {Ressler}, {Cutri}, {Jarrett}, {Kirkpatrick}, {Padgett}, {McMillan}, {Skrutskie}, {Stanford}, {Cohen}, {Walker}, {Mather}, {Leisawitz}, {Gautier}, {McLean}, {Benford}, {Lonsdale}, {Blain}, {Mendez}, {Irace}, {Duval}, {Liu}, {Royer}, {Heinrichsen}, {Howard}, {Shannon}, {Kendall}, {Walsh}, {Larsen}, {Cardon}, {Schick}, {Schwalm}, {Abid}, {Fabinsky}, {Naes}, \& {Tsai}}]{wise}
{Wright}, E.~L., {Eisenhardt}, P. R.~M., {Mainzer}, A.~K., {et~al.} 2010, \aj, 140, 1868

\end{thebibliography}

%-------------------------------------------------------------------

\clearpage
\onecolumn

\begin{appendix}
\section{Additional table}

\LTcapwidth=\textwidth
\begin{longtable}{llrllrc}
\caption{\label{Tab:SNR_list}SNRs considered in this work: for each of them, it is reported the radio morphology classification, information on their remnant (extracted from the University of Manitoba catalogue$^{1}$) and the cluster ID from the reference experiment B3. The last column reports alternative ID. SNR candidates found with MOST and included in the sample are reported with $\ddagger$.}\\

\hline\hline
SNR Name        & RA (J2000) & Dec. (J2000) & Type          & Remnant       & Cluster & ID\_alt                \\ 
       &(hh:mm:ss) & (dd:mm:ss) &                            &        &  &                \\ 
\hline
\endfirsthead
\caption{continued.}\\
\hline\hline
SNR Name        & RA (J2000) & Dec. (J2000) & Type          & Remnant       & Cluster & ID\_alt                \\ 
       &(hh:mm:ss) & (dd:mm:ss) &                            &        &  &                \\ 
\hline       
\endhead
\hline       
\endfoot
G003.7-00.2 & 17:55:26             & -25:50:00             & shell                          & eje           & 0       &                        \\
G011.1-00.7 & 18:12:46             & -19:38:00             & shell                          & eje           & 0       &                        \\
G021.6-00.8 & 18:33:40             & -10:25:00             & shell                          & eje           & 0       &                        \\
G031.5-00.6 & 18:51:10             & -01:31:00             & shell                          & eje?          & 0       &                        \\
G032.1-00.9 & 18:53:10             & -01:08:00             & thermal composite              & eje?          & 0       &                        \\
G033.2-00.6 & 18:53:50             & -00:02:00             & shell                          & eje           & 0       &                        \\
G036.6-00.7 & 19:00:35             & 02:56:00              & shell                          & eje?          & 0       &                        \\
G299.6-00.5 & 12:21:45             & -63:09:00             & shell                          & eje           & 0       &                        \\
G302.3+00.7 & 12:45:55             & -62:08:00             & shell                          & eje           & 0       &                        \\
G310.6-00.3 & 13:58:00             & -62:09:00             & shell                          & eje           & 0       & Kes 20B                \\
G312.4-00.4 & 14:13:00             & -61:44:00             & shell                          & eje,PSR,PWN?  & 0       &                        \\
G341.9-00.3 & 16:55:01             & -44:01:00             & shell                          & eje           & 0       &                        \\
G342.0-00.2 & 16:54:50             & -43:53:00             & shell                          & eje           & 0       &                        \\
G343.1-00.7 & 17:00:25             & -43:14:00             & shell                          & eje           & 0       &                        \\
G346.6-00.2 & 17:10:19             & -40:11:00             & thermal composite              & eje           & 0       &                        \\
G351.2+00.1 & 17:22:27             & -36:11:00             & plerionic composite            & eje           & 0       &                        \\
G355.6-00.0 & 17:35:16             & -32:38:00             & thermal composite              & eje           & 0       &                        \\
G001.9+00.3 & 17:48:45             & -27:10:00             & shell                          & eje           & 1       &                        \\
G006.1+00.5 & 17:57:29             & -23:25:00             & shell                          & eje           & 1       &                        \\
G009.8+00.6 & 18:05:08             & -20:14:00             & shell                          & eje           & 1       &                        \\
G012.2+00.3 & 18:11:17             & -18:10:00             & shell                          & eje           & 1       &                        \\
G015.9+00.2 & 18:18:52             & -15:02:00             & shell                          & eje,NS        & 1       &                        \\
G042.8+00.6 & 19:07:20             & 09:05:00              & shell                          & eje,NS?,PSR?  & 1       &                        \\
G318.9+00.4 & 14:58:30             & -58:29:00             & plerionic composite            & eje?,PWN?     & 1       &                        \\
G322.1+00.0 & 15:20:49             & -57:10:00             & shell                          & eje,NS        & 1       & Circinus X-1           \\
G327.4+01.0 & 15:46:48             & -53:20:00             & shell                          & eje           & 1       &                        \\
G335.2+00.1 & 16:27:45             & -48:47:00             & shell                          & eje,PSR       & 1       &                        \\
G336.7+00.5 & 16:32:11             & -47:19:00             & shell                          & eje           & 1       &                        \\
G337.3+01.0 & 16:32:39             & -46:36:00             & shell                          & eje           & 1       & Kes 40                 \\
G338.1+00.4 & 16:37:59             & -46:24:00             & shell                          & eje           & 1       &                        \\
G006.1+01.2 & 17:54:55             & -23:05:00             & filled-centre                  & ?             & 2       &                        \\
G012.8-00.0 & 18:13:37             & -17:49:00             & plerionic composite            & eje,PSR?,PWN? & 2       & W33                    \\
G016.0-00.5 & 18:21:56             & -15:14:00             & shell                          & eje           & 2       &                        \\
G030.7+01.0 & 18:44:00             & -01:32:00             & shell                          & eje?          & 2       &    \\
G321.9-01.1 & 15:23:45             & -58:13:00             & shell                          & eje           & 2       &                        \\
G322.9-0.0$\ddagger$  & 15:25:41             & -56:46:16             &      unknown                          &               & 2       &           \\
G348.8+1.1$\ddagger$  & 17:11:29             & -37:35:39             &      unknown                          &               & 2       &          \\
G006.4-00.1 & 18:00:30             & -23:26:00             & thermal composite              & eje           & 3       & W28                    \\
G012.0-00.1 & 18:12:11             & -18:37:00             & shell                          & eje           & 3       &                        \\
G028.6-00.1 & 18:43:55             & -03:53:00             & shell                          & eje           & 3       &                        \\
G049.2-00.7 & 19:23:50             & 14:06:00              & thermal composite              & eje?,NS?,PWN? & 3       & W51C                   \\
G308.8-00.1 & 13:42:30             & -62:23:00             & plerionic composite            & eje?,PSR,PWN  & 3       &                        \\
G315.4-00.3 & 14:35:55             & -60:36:00             & unknown                        & ?             & 3       &                        \\
G332.4+00.1 & 16:15:20             & -50:42:00             & shell                          & eje           & 3       & Kes 32, MSH 16-51      \\
G337.0-00.1 & 16:35:57             & -47:36:00             & shell                          & eje,NS?       & 3       & (CTB 33)               \\
G344.7-00.1 & 17:03:51             & -41:42:00             & thermal composite              & eje,NS?       & 3       &                        \\
G347.3-00.5 & 17:13:28             & -39:49:48             & shell                          & eje,NS        & 3       & RX J1713.7-3946        \\
G348.5-00.0 & 17:15:26             & -38:28:00             & shell                          & eje?          & 3       &                        \\
G348.5+00.1 & 17:14:40             & -38:32:00             & thermal composite              & eje,PWN?      & 3       & CTB 37A                \\
G349.2-00.1 & 17:17:15             & -38:04:00             & shell                          & eje,PSR       & 3       &                        \\
G353.6-00.7 & 17:32:00             & -34:44:00             & shell                          & eje,NS?,PSR?  & 3       & J1731-347              \\
G357.7-00.1 & 17:40:29             & -30:58:00             & thermal composite              & eje?          & 3       & The Tornado, MSH 17-39 \\
G005.5+00.3 & 17:57:04             & -24:00:00             & shell                          & eje           & 4       &                        \\
G008.7-00.1 & 18:05:30             & -21:26:00             & plerionic composite & eje,PSR?,PWN?                         & 4       & (W30)                  \\
G009.9-00.8 & 18:10:41             & -20:43:00             & shell                          & eje           & 4       &                        \\
G012.7-00.0 & 18:13:19             & -17:54:00             & shell                          & eje           & 4       &                        \\
G017.4-00.1 & 18:23:08             & -13:46:00             & shell                          & eje           & 4       &                        \\
G018.8+00.3 & 18:23:58             & -12:23:00             & shell                          & eje           & 4       & Kes 67                 \\
G023.3-00.3 & 18:34:45             & -08:48:00             & shell                          & eje,PSR,PWN?  & 4       & W41                    \\
G032.8-00.1 & 18:51:25             & -00:08:00             & shell                          & eje           & 4       & Kes 78                 \\
G034.7-00.4 & 18:56:00             & 01:22:00              & plerionic composite & eje,PSR,PWN   & 4       & W44                    \\
G042.0-00.1 & 19:08:10             & 08:00:00              & shell                          & eje?          & 4       &                        \\
G045.7-00.4 & 19:16:25             & 11:09:00              & shell                          & eje           & 4       &                        \\
G286.5-01.2 & 10:35:40             & -59:42:00             & shell                          & eje?          & 4       &                        \\
G290.1-00.8 & 11:03:05             & -60:56:00             & thermal composite              & eje,PSR?,PWN? & 4       & MSH 11-61A             \\
G296.7-00.9 & 11:55:31             & -63:07:08             & shell                          & eje           & 4       &                        \\
G308.7+0.0$\ddagger$  & 13:41:30             & -62:15:06             &   unknown           &               & 4       &                        \\
G310.8-00.4 & 14:00:00             & -62:17:00             & shell                          & eje           & 4       & Kes 20A                \\
G321.9-00.3 & 15:20:40             & -57:34:00             & shell                          & eje,PSR?      & 4       &                        \\
G323.5+00.1 & 15:28:42             & -56:21:00             & shell                          & eje           & 4       &                        \\
G324.1+0.1$\ddagger$  & 15:32:32             & -56:03:08             &        unknown                        &               & 4       &           \\
G327.4+00.4 & 15:48:20             & -53:49:00             & thermal composite              & eje           & 4       & Kes 27                 \\
G329.7+00.4 & 16:01:20             & -52:18:00             & shell                          & eje           & 4       &                        \\
G332.0+00.2 & 16:13:17             & -50:53:00             & shell                          & eje           & 4       &                        \\
G338.3-00.0 & 16:41:00             & -46:34:00             & plerionic composite            & eje,PSR,PWN   & 4       &                        \\
G012.5+00.2 & 18:12:14             & -17:55:00             & plerionic composite            & eje?          & 5       &                        \\
G013.5+00.2 & 18:14:14             & -17:12:00             & shell                          & eje           & 5       &                        \\
G315.9-00.0 & 14:38:25             & -60:11:00             & plerionic composite            & eje,PSR,PWN   & 5       &                        \\
G320.6-0.9$\ddagger$  & 15:15:03             & -58:46:55             &   unknown                             &               & 5       &           \\
G337.2+00.1 & 16:35:55             & -47:20:00             & plerionic composite            & eje,PWN?      & 5       &                        \\
G341.2+00.9 & 16:47:35             & -43:47:00             & plerionic composite            & eje,PSR,PWN   & 5       &                        \\
G011.1-01.0 & 18:14:03             & -19:46:00             & shell                          & eje           & 6       &                        \\
G011.1+00.1 & 18:09:47             & -19:12:00             & plerionic composite            & eje,PSR,PWN   & 6       &                        \\
G032.4+00.1 & 18:50:05             & -00:25:00             & shell                          & eje           & 6       &                        \\
G289.7-00.3 & 11:01:15             & -60:18:00             & shell                          & eje           & 6       &                        \\
G332.4-00.4 & 16:17:33             & -51:02:00             & shell                          & eje,NS?       & 6       & RCW 103                \\
G340.6+00.3 & 16:47:41             & -44:34:00             & shell                          & eje           & 6       &                        \\
G357.7+00.3 & 17:38:35             & -30:44:00             & shell                          & eje           & 6       & the Square             \\
G011.4-00.1 & 18:10:47             & -19:05:00             & shell                          & eje           & 7       &                        \\
G029.7-00.3 & 18:46:25             & -02:59:00             & plerionic composite            & eje,PSR,PWN   & 7       & Kes 75                 \\
G319.9-0.7$\ddagger$   & 15:09:14             & -58:54:26             &  unknown                              &               & 7       &           \\
G323.2-1.0$\ddagger$  & 15:31:41             & -57:23:55             &       unknown                         &               & 7       &            \\
G327.1-01.1 & 15:54:25             & -55:09:00             & plerionic composite            & eje,PWN       & 7       &                        \\
G328.4+00.2 & 15:55:30             & -53:17:00             & filled-centre                  & NS,PWN        & 7       & MSH 15-57              \\
G339.6-0.6$\ddagger$  & 16:13:13             & -51:11:38             &    unknown                            &               & 7       &            \\
G003.8+00.3 & 17:52:55             & -25:28:00             & shell                          & eje           & -1      &                        \\
G006.5-00.4 & 18:02:11             & -23:34:00             & shell                          & eje           & -1      &                        \\
G007.0-00.1 & 18:01:50             & -22:54:00             & shell                          & eje           & -1      &                        \\
G007.2+00.2 & 18:01:07             & -22:38:00             & shell                          & eje           & -1      &                        \\
G008.3-00.0 & 18:04:34             & -21:49:00             & shell                          & eje           & -1      &                        \\
G008.9+00.4 & 18:03:58             & -21:03:00             & shell                          & eje           & -1      &                        \\
G009.7-00.0 & 18:07:22             & -20:35:00             & shell                          & eje           & -1      &                        \\
G010.5-00.0 & 18:09:08             & -19:47:00             & shell                          & eje           & -1      &                        \\
G011.0-00.0 & 18:10:04             & -19:25:00             & shell                          & eje,PWN?      & -1      &                        \\
G011.2-00.3 & 18:11:29             & -19:25:25             & plerionic composite            & eje,PSR,PWN   & -1      &                        \\
G011.8-00.2 & 18:12:25             & -18:44:00             & shell                          & eje           & -1      &                        \\
G014.1-00.1 & 18:16:40             & -16:41:00             & shell                          & eje           & -1      &                        \\
G014.3+00.1 & 18:15:58             & -16:27:00             & shell                          & eje           & -1      &                        \\
G015.4+00.1 & 18:18:02             & -15:27:00             & plerionic composite            & eje,PWN?      & -1      &                        \\
G016.4-00.5 & 18:22:38             & -14:55:00             & shell                          & eje           & -1      &                        \\
G017.0-00.0 & 18:21:57             & -14:08:00             & shell                          & eje           & -1      &                        \\
G018.1-00.1 & 18:24:34             & -13:11:00             & shell                          & eje           & -1      &                        \\
G018.6-00.2 & 18:25:55             & -12:50:00             & shell                          & eje           & -1      &                        \\
G019.1+00.2 & 18:24:56             & -12:07:00             & shell                          & eje           & -1      &                        \\
G020.0-00.2 & 18:28:07             & -11:35:00             & filled-centre                  & PWN?          & -1      &                        \\
G020.4+00.1 & 18:27:51             & -11:00:00             & shell                          & eje           & -1      &                        \\
G021.0-00.4 & 18:31:12             & -10:47:00             & shell                          & eje           & -1      &                        \\
G021.5-00.1 & 18:30:50             & -10:09:00             & shell                          & eje           & -1      &                        \\
G021.8-00.6 & 18:32:45             & -10:08:00             & thermal composite              & eje           & -1      & Kes 69                 \\
G022.7-00.2 & 18:33:15             & -09:13:00             & shell                          & eje?          & -1      &                        \\
G023.6+00.3 & 18:33:03             & -08:13:00             & unknown                        & eje?          & -1      &                        \\
G024.7-00.6 & 18:38:43             & -07:32:00             & shell                          & eje?          & -1      &                        \\
G027.4+00.0 & 18:41:19             & -04:56:00             & shell                          & eje,PSR       & -1      & Kes 73                 \\
G027.8+00.6 & 18:39:50             & -04:24:00             & filled-centre                  & PWN?          & -1      &                        \\
G029.6+00.1 & 18:44:52             & -02:57:00             & shell                          & eje,PSR?      & -1      &                        \\
G031.9+00.0 & 18:49:25             & -00:55:00             & thermal composite              & eje           & -1      & Kes 77                 \\
G033.6+00.1 & 18:52:39             & 00:40:20              & thermal composite              & eje,PSR       & -1      & Kes 79                 \\
G035.6-00.4 & 18:57:55             & 02:13:00              & shell                          & eje,PSR?      & -1      &                        \\
G039.2-00.3 & 19:04:08             & 05:28:00              & plerionic composite            & eje,PWN       & -1      &                        \\
G040.5-00.5 & 19:07:10             & 06:31:00              & shell                          & eje,PSR,PWN?  & -1      &                        \\
G041.1-00.3 & 19:07:34             & 07:08:00              & thermal composite              & eje           & -1      & 3C397                  \\
G041.5+00.4 & 19:05:50             & 07:46:00              & shell                          & eje?          & -1      &                        \\
G043.3-00.2 & 19:11:08             & 09:06:00              & thermal composite              & eje           & -1      & W49B, (3C398)          \\
G046.8-00.3 & 19:18:10             & 12:09:00              & shell                          & eje           & -1      & (HC30)                 \\
G054.1+00.3 & 19:30:31             & 18:52:00              & plerionic composite            & eje?,PSR,PWN  & -1      &                        \\
G054.4-00.3 & 19:33:20             & 18:56:00              & shell                          & eje,PSR?      & -1      & (HC40)                 \\
G055.0+00.3 & 19:32:00             & 19:50:00              & shell                          & eje,PSR?      & -1      &                        \\
G059.5+00.1 & 19:42:33             & 23:35:00              & shell                          & eje           & -1      &                        \\
G059.8+01.2 & 19:38:55             & 24:19:00              & unknown                        & eje?          & -1      &                        \\
G284.3-01.8 & 10:18:15             & -59:00:00             & shell                          & eje,NS        & -1      & MSH 10-53              \\
G291.0-00.1 & 11:11:54             & -60:38:00             & plerionic composite            & eje,PSR?,PWN  & -1      & (MSH 11-62)            \\
G292.2-00.5 & 11:19:20             & -61:28:00             & plerionic composite & eje,PSR,PWN   & -1      &                        \\
G294.1-00.0 & 11:36:10             & -61:38:00             & shell                          & eje           & -1      &                        \\
G296.1-00.5 & 11:51:10             & -62:34:00             & shell                          & eje           & -1      &                        \\
G296.8-00.3 & 11:58:30             & -62:35:00             & shell                          & eje,PSR?      & -1      & 1156-62                \\
G298.5-00.3 & 12:12:40             & -62:52:00             & unknown                        & eje?          & -1      &                        \\
G298.6-00.0 & 12:13:41             & -62:37:00             & thermal composite              & eje           & -1      &                        \\
G301.4-01.0 & 12:37:55             & -63:49:00             & shell                          & eje           & -1      &                        \\
G304.6+00.1 & 13:05:59             & -62:42:00             & thermal composite              & eje           & -1      & Kes 17                 \\
G306.3-00.9 & 13:21:50             & -63:34:00             & shell                          & eje           & -1      &                        \\
G308.1-00.7 & 13:37:37             & -63:04:00             & shell                          & eje           & -1      &                        \\
G309.2-00.6 & 13:46:31             & -62:54:00             & shell                          & eje           & -1      &                        \\
G309.8+00.0 & 13:50:30             & -62:05:00             & shell                          & eje           & -1      &                        \\
G316.3-00.0 & 14:41:30             & -60:00:00             & shell                          & eje           & -1      & (MSH 14-57)            \\
G317.3-00.2 & 14:49:40             & -59:46:00             & shell                          & eje           & -1      &                        \\
G318.2+00.1 & 14:54:50             & -59:04:00             & shell                          & eje           & -1      &                        \\
G322.5-00.1 & 15:23:23             & -57:06:00             & plerionic composite            & eje,PWN?      & -1      &                        \\
G322.7+0.1$\ddagger$  & 15:23:56             & -56:48:41             &      unknown                          &               & -1      &           \\
G325.0-0.3$\ddagger$  & 15:39:13             & -55:49:45             &        unknown                        &               & -1      &            \\
G327.2-00.1 & 15:50:55             & -54:18:00             & shell                          & eje,PSR       & -1      &                        \\
G330.2+01.0 & 16:01:06             & -51:34:00             & shell                          & eje,NS        & -1      &                        \\
G331.8-0.0$\ddagger$  & 16:13:13             & -51:11:38             &      unknown                          &               & -1      &           \\
G337.2-00.7 & 16:39:28             & -47:51:00             & shell                          & eje           & -1      &                        \\
G337.8-00.1 & 16:39:01             & -46:59:00             & thermal composite              & eje           & -1      & Kes 41                 \\
G338.5+00.1 & 16:41:09             & -46:19:00             & unknown                        & eje?          & -1      &                        \\
G340.4+00.4 & 16:46:31             & -44:39:00             & shell                          & eje           & -1      &                        \\
G342.1+00.9 & 16:50:43             & -43:04:00             & shell                          & eje           & -1      &                        \\
G345.1-0.2$\ddagger$  & 17:05:21             & -41:26:04             &    unknown                            &               & -1      &            \\
G345.1+0.2$\ddagger$  & 17:03:40             & -41:05:11             &   unknown                             &               & -1      &           \\
G345.7-00.2 & 17:07:20             & -40:53:00             & shell                          & eje           & -1      &                        \\
G350.1-00.3 & 17:21:00             & -37:24:00             & shell                          & eje,NS?       & -1      &                        \\
G351.7+00.8 & 17:21:00             & -35:27:00             & shell                          & eje           & -1      &                        \\
G351.9-00.9 & 17:28:52             & -36:16:00             & shell                          & eje           & -1      &                        \\
G352.7-00.1 & 17:27:40             & -35:07:00             & thermal composite              & eje           & -1      &                        \\
G354.1+00.1 & 17:30:28             & -33:46:00             & plerionic composite            & eje?,PSR?     & -1      &                        \\
G354.8-00.8 & 17:36:00             & -33:42:00             & shell                          & eje           & -1      &                        \\
G355.4+00.7 & 17:31:20             & -32:26:00             & shell                          & eje           & -1      &                        \\
G356.3-00.3 & 17:37:56             & -32:16:00             & shell                          & eje           & -1      &                        \\

\hline

\end{longtable}
\tablebib{
$^1$  \citealt{Ferrand2012}}

\section{Impact of masking compact sources}\label{app:masked-sources}

 The automated binary masks produced by \textsc{caesar} to remove compact sources during the preprocessing of the input images are not perfect. In some cases, small but noticeable residuals appear around the masked regions,  particularly masking very bright sources or sources that deviate from a Gaussian profile. These residuals may `contaminate' the embeddings learned by the CAE, thus reappearing in the reconstructed images as diffuse bright blobs or artefacts (see, e.g., some features in the examples from cluster 3 in Figure \ref{fig:best-cluster}).

\begin{figure}[!h]
    \centering
    \includegraphics[width=\textwidth]{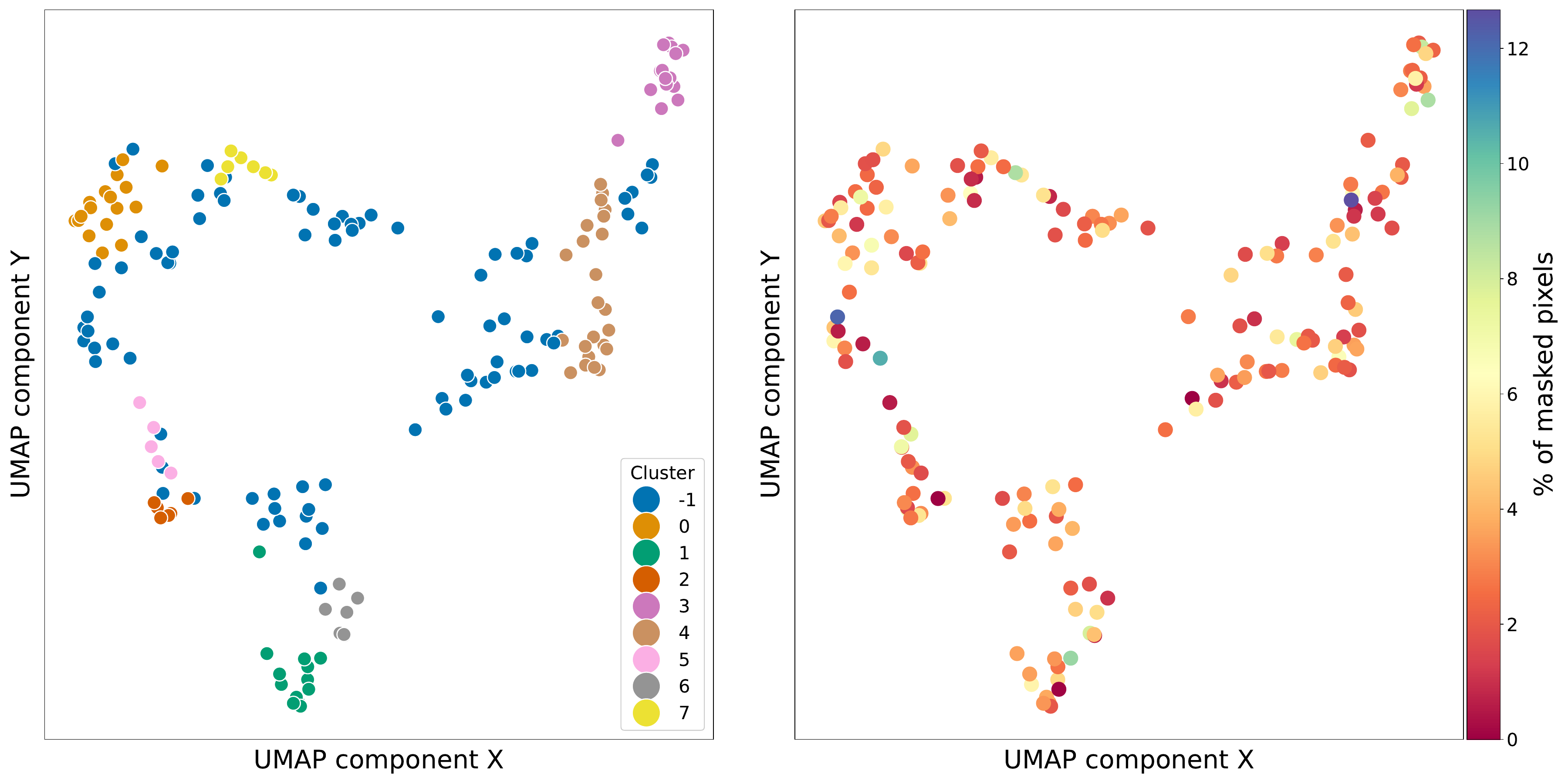}
    \caption{Impact of source masking in clustering results. Left: Visualisation of DBSCAN clustering of experiment B3 SNR embeddings (same as in Figure \ref{fig:best-cluster}, panel a). Right: Same experiment, but colour-coded based on the amount of masked pixels in the original input images.}
    \label{fig:blank-pixels}
\end{figure}

To measure the extent to which these artefacts in the input images may influence the clustering results, we computed the amount of masked pixels per input image. We only consider those pixels masked because of the presence of compact sources, meaning that pixels outside the circular region of interest applied to each source are not included in the count. In Figure \ref{fig:blank-pixels} we present a side-by-side comparison of the representative experiment B3 (see Section \ref{sec:discussion}), comparing the assigned cluster labels with a colour scale that represents the amount of masked pixels per instance. From the plot, it is evident that there is not a clear trend: the amount of masked pixels varies from source to source within a given cluster, and all clusters have mean and median values of $\sim$2--4\%. The inter-cluster standard deviation is $\sim$0.7, showing little differences among clusters. This indicates that, even if we cannot completely rule out an influence, masking out compact sources is not the dominant factor affecting the clustering results.

\end{appendix}

%\begin{thebibliography}{}

%\end{thebibliography}

\end{document}